\documentclass[aps, preprint, nofootinbib,preprintnumbers,eqsecnum]{revtex4}

\usepackage[dvips]{color}
\usepackage[normalem]{ulem}
\usepackage{amsmath}
\usepackage{amssymb}
\usepackage{amscd}
\usepackage{enumerate}
\usepackage{amsfonts}
\usepackage{epsfig}
\usepackage{yfonts}

\usepackage{dsfont}

\usepackage{graphicx}
\usepackage{bm}

\definecolor{davecolor}{rgb}{0.95,  0.5,  0.2}

\def\({\left(}
\def\){\right)}
\def\[{\left[}
\def\]{\right]}
\def\<{\langle}
\def\>{\rangle}

\def\Tr{\mathop{\rm Tr}}

\newcommand\ket[1]{\ensuremath{\lvert{#1}\rangle}}
\newcommand\bra[1]{\ensuremath{\langle{#1}\rvert}}

\newcommand{\be}{\begin{equation}}
\newcommand{\ee}{\end{equation}}
\newcommand{\bea}{\begin{eqnarray}}
\newcommand{\eea}{\end{eqnarray}}
\newcommand{\bwt}{\begin{widetext}}
\newcommand{\ewt}{\end{widetext}}

\newcommand{\bi}{\begin{itemize}}
\newcommand{\ei}{\end{itemize}}
\newcommand{\ben}{\begin{enumerate}}
\newcommand{\een}{\end{enumerate}}
\newcommand{\bca}{\begin{cases}}
\newcommand{\eca}{\end{cases}}
\newcommand{\bln}{\begin{align}}
\newcommand{\eln}{\end{align}}
\newcommand{\bst}{\begin{split}}
\newcommand{\est}{\end{split}}

\renewcommand{\Im}{\textrm{Im}\,}

\begin{document}

\title{The Entanglement Renyi Entropies of Disjoint Intervals in AdS/CFT}

\author{Thomas Faulkner}
\affiliation{Institute for Advanced Study, Princeton, NJ, 08540}
\email{faulkner@ias.edu}

\begin{abstract}

We study entanglement Renyi entropies (EREs) of $1+1$ dimensional CFTs with classical
gravity duals. Using the replica trick the EREs can be related to a partition function of $n$ copies of the CFT glued together in a particular way along the intervals. 
In the case of two intervals this procedure defines a genus $n-1$ surface and our goal is to find smooth three dimensional gravitational solutions with this surface living at the boundary. We find two families of handlebody solutions labelled by the replica index $n$. These particular bulk solutions are distinguished by the
fact that they do not spontaneously break the replica symmetries of the boundary  surface.
We show that the regularized classical action of these solutions is given in terms of
a simple numerical prescription.  
If we assume that they give the dominant contribution to the gravity partition
function we can relate this classical action to the EREs at leading order in $G_N$. We argue that the prescription can be formulated for non-integer $n$. Upon taking the limit $n \rightarrow 1$ 
the classical action reproduces the predictions of the Ryu-Takayanagi formula for the entanglement entropy.
\end{abstract} 

\today

\maketitle

\tableofcontents

\section{Introduction}

Entanglement entropy (EE) is a powerful observable 
for many-body quantum systems. This is especially so when defined with respect
to the reduced density matrix associated to a spatial subregion $\mathcal{A}$ of the
full system \cite{Holzhey:1994we,Srednicki:1993im}. 
EE then detects spatial quantum correlations in a fixed  many-body state. 
One simple reason for the appeal of EE is the universal
nature of its definition allowing for model independent characterizations
of many-body phases.  To list a few applications: EE has been used as an order parameter to distinguish trivially gapped phases from those with topological degrees of freedom
\cite{Kitaev:2005dm,wen}, as a c-function
on CFTs in two and three dimensions \cite{Casini:2006es,Casini:2012ei,Liu:2012eea} and as a measure of thermalization in non-equilibrium situations \cite{cctherm}.

Unfortunately EEs are rather hard to compute theoretically even for free theories.
Techniques for CFTs are available \cite{Casini:2011kv,Calabrese:2004eu,Calabrese:2009tt} and give results for fairly simple spatial regions $\mathcal{A}$. However a more general understanding of EE in 
QFT is lacking.

Surprisingly there is a simple formula for computing EE in AdS/CFT
given by a prescription of Ryu and Takayanagi (RT) \cite{Ryu:2006bv,Ryu:2006ef,Nishioka:2009un} involving the area of minimal surfaces.  The formula applies to
quantum field theories with dual classical Einstein gravity descriptions. Some higher derivative corrections have been attempted, see for example \cite{Hung:2011xb}, while bulk quantum corrections are unknown. The status of the formula remains as a further conjecture above and beyond the  usual rules of the Maldacena conjecture \cite{Maldacena:1997re,Witten:1998qj,Gubser:1998bc}. In principle one should be able to derive it using just these  rules, however the attempt in \cite{Fursaev:2006ih} failed as was emphasized in \cite{Headrick:2010zt}.  In particular a derivation would forge the way to understanding bulk quantum and classical corrections to the formula.

\begin{figure}[h!]
\centering
\includegraphics[scale=.5]{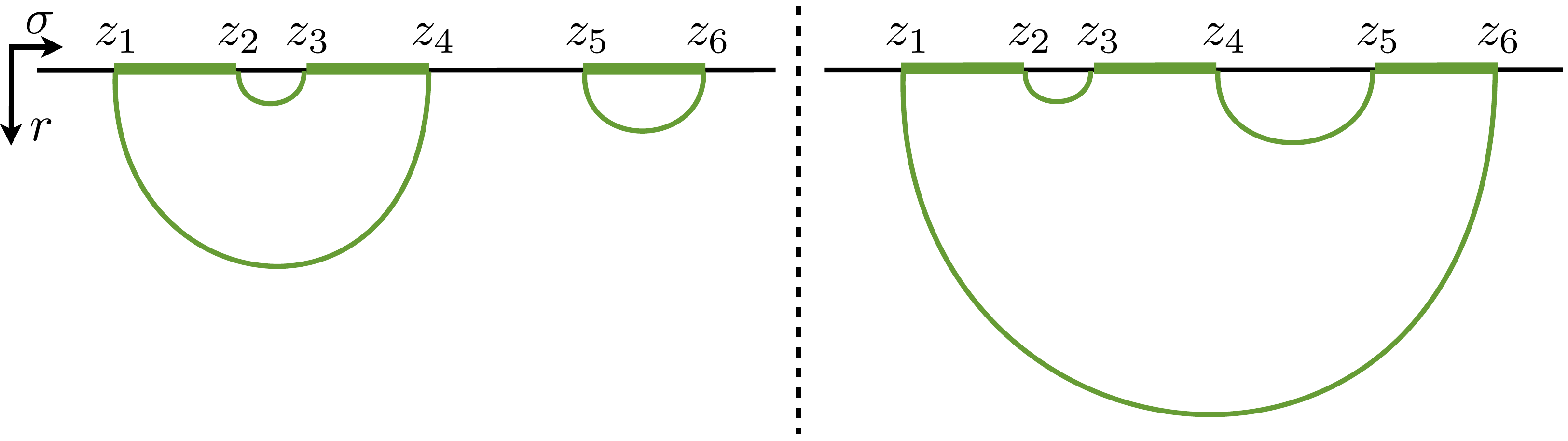}
\caption{The RT prescription for computing the EE in 1+1 dimensional CFTs for 3 disjoint intervals.
The CFT spatial direction is $\sigma$ and $r$ is the radial direction of the dual $AdS_3$.  The minimal surfaces are simply geodesics connecting the ends of the intervals.
The sum of the regularized lengths of these geodesics computes the EE.
There is more than one minimal set of such geodesics and one is instructed
to find the global minimum. We have shown only 2 cases out of a total of 5. \label{fig:rt}}
\end{figure}

The focus of this paper will be $1+1$ CFTs  where the sub-region of interest
$\mathcal{A}$ is the union of a set of intervals along the spatial axis \cite{Calabrese:2009ez,Calabrese:2004eu} and we consider only the vacuum state of the CFT. 
The RT prediction for this case was discussed
in \cite{Headrick:2010zt,Swingle:2010jz} and involves the lengths of bulk geodesics
which we summarize in Figure~\ref{fig:rt}.
We will attempt to prove the RT formula for this case using the replica trick. This trick involves calculating the Entanglement Renyi Entropies (ERE) as an intermediate step
\be
\label{ere}
S_n = - \frac{1}{n-1} \ln {\rm Tr}_{\mathcal{A}} (\rho_{\mathcal A})^n
\ee
where $\rho_{\mathcal A}$ is the reduced density matrix in the vacuum of the CFT 
for the Hilbert space
associated to the intervals $\mathcal{A}$. The EREs are defined for integer $n\geq2$ 
and can be calculated by the partition function of the CFT on a 
surface $\mathcal{M}$ of genus $n-1$. 
\emph{Assuming} one can analytically continue the partition function to non-integer $n$ then the limit $n \rightarrow 1$ gives the von Neumann entropy expression for the EE.

This paper was inspired by some of the results of  Headrick in \cite{Headrick:2010zt}
where the ERE for two intervals and $n=2$  was found for CFTs with gravitational duals.
We  attempt to generalize Headrick's results by finding the gravity solutions  which are needed to compute the EREs holographically for $n>2$. 
We seek handlebody solutions whose conformal boundary is
the genus $n-1$ surface $\mathcal{M}$.
To generate such solutions we need to represent $\mathcal{M}$ in terms of its so called Schottky uniformization. This  representation of $\mathcal{M}$ can be roughly described as
a connected domain in the complex plane  with certain identifications on the boundaries of the domain.
Schottky uniformization allows us to find the bulk
handlebody solution by extending  
the domain boundaries and identifications into the bulk radial direction in a particular way.

Actually there is an infinite set of such gravitational solutions. At finite  Newton's constant $G_N$ one expects all of these to contribute to the partition function as
\be
\label{sadsum}
Z_\mathcal{M} = \sum_{\gamma} \exp( -S_{gr}^\gamma  + \mathcal{O}(G_N^0))
\ee
where $S_{gr}^\gamma \propto G_N^{-1} $ is the gravitational action for the classical solution 
labelled by $\gamma$. However in the classical limit where $G_N \rightarrow 0$ only  the least action solution will dominate and we only need to find this one.
In this paper we show that one can  easily construct  a small finite subset of the infinite set of solutions that contribute to \eqref{sadsum}.

Interestingly the solutions we can construct in this way have the property that one can
formulate a simple numerical problem which computes their gravitational action. 
The answer can then be found  numerically for integer $n \geq 2$.
This formulation can be continued in the replica index $n$ to non-integers. This
is true despite the fact that the bulk solutions no longer make any sense. The limit
$n \rightarrow 1$ can be studied exactly and the actions computed in this 
way reproduce the Ryu-Takayanagi formula for the EE involving the lengths of  bulk geodesics.

Unfortunately since it is more difficult to construct the missing solutions in \eqref{sadsum} to check that they are all subdominant we are left only with a partial result. The gravitational actions we
compute via the numerical prescription can only be related to the EREs if we assume they are in fact the dominant ones. If one could show that this assumption is correct then we could
compute the EREs and prove the RT formula. 
A simple way to characterize
the missing handlebody saddles is by the fact that the bulk solution breaks some of the symmetries of the boundary manifold including for example the replica symmetry which interchanges the different replicas.\footnote{
There are also non-handlebody solutions which are usually assumed to be subdominant
since they would be pathological from an AdS/CFT point of view. We come back to these 
as well as the replica breaking saddles in the discussion. } This would be an interesting phenomena if it were to happen, and the investigation
of this possibility is left to the future. 

Although we will discuss some results for multiple intervals most of
the discussion will be for the case of 2 intervals. We expect our results to generalize
to multiple intervals.

Our results match the calculations of a complementary paper \cite{hartman} which takes the CFT perspective to this problem. CFTs with large $c$ and a a small number of low dimension
primary operators were considered. The arguments in \cite{hartman} are based on semiclassical conformal blocks.  We comment more on this paper in the discussion.

The paper is organized as follows. 
In Section~II we introduce the replica trick which tells us to compute
the partition function of a certain genus $n-1$ Riemann surface $\mathcal{M}$ the properties
of which we also discuss here. 
In Section~III we give a numerical prescription
for computing EREs in $1+1$ CFTs with a classical gravity description.
This prescription remains a conjecture since we could not rule out the possibility
of other saddles being dominant. However since the end result we found is
rather simple it is useful to present this  before delving into the details of
its derivation.  We subsequently show that these saddles reproduce the RT prescription
and reproduce other known results in the literature. In Section~IV we discuss the essential
ideas behind the program of Schottky uniformization. 
In the Section~V we gives details of the bulk solutions that we find.
In Section~VI we compute the bulk action in a few ways and relate the answer
to the prescription given earlier on. We end with a discussion. 
There are several appendices with details.

\section{The replica trick and the Riemann surface}

We are interested in computing the ERE for a spatial region $\mathcal{A}$ - the set of $N$ intervals:
\be
\mathcal{A} = [z_1, z_{2}] \cup  [z_3, z_4 ] \ldots \cup [z_{2N-1},z_{2N}]
\ee 
where the $z_i$ are cyclicly ordered. 
The Hilbert space factors locally: $\mathcal{H} = \mathcal{H}_{\mathcal A}
\otimes \mathcal{H}_{\mathcal{A}^c}$ where $\mathcal{A}^c$ is the complement region to the above intervals. The EE  in the vacuum state is defined by:
\be
\rho_{\mathcal A}  = {\rm Tr}_{\mathcal{A}^c}  \ket{0} \bra{0} \quad \rightarrow \quad S_{EE}  = -  {\rm Tr}_{\mathcal A} \left( \rho_{\mathcal A} \log \rho_{\mathcal A} \right)
\ee
and the ERE generalizations were given in \eqref{ere} such that
$\lim_{n \rightarrow 1} S_n = S_{EE}$. The replica trick allows one to formulate $\Tr (\rho_{\mathcal A})^n$ as a partition function of the theory on a particular manifold. The arguments are standard and can be found for example in the review \cite{Headrick:2012fk}. 
For each of the $n$ factors of $\rho_{\mathcal A}$ one introduces a Euclidean path integral
on the complex $z$-plane with certain boundary conditions on the $z$ real axis. The
trace and sum over intermediate states then glues  together these $n$ copies of the $z$-plane
along the intervals in $\mathcal{A}$ in a particular way.
The result is a Euclidean path integral on an $n$-sheeted Riemann surface
or branched covering defined by:
\be
\label{rs}
\mathcal{M} : \qquad y^n = \prod_{i=1\ldots N} \frac{(z-z_{2i-1})}{(z-z_{2i})}
\ee
with the entanglement region $\mathcal{A}$ lying on the real $z$ axis.\footnote{ Some properties of this surface for
two intervals $N=2$ are summarized in Appendix~A.}
The genus of this surface is $(N-1)(n-1)$. Beyond this
specifying \eqref{rs} only tells us the complex structure of the surface, however to compute the CFT partition function we also need to give a particular metric in the fixed conformal class. We take
this to be the original metric that the CFT lives on:
\be
\label{ds2}
ds^2 = d z d \bar{z}
\ee
On the branched covering this metric necessarily has conical excess singularities at the branch points. These can be resolved by cutting out a region $\epsilon$ from the branch points and
replacing the singular metric with a smooth one.  The details of this procedure are 
standard and given in Appendix~\ref{sec:cut}.  The Euclidean path integral on $\mathcal{M}$ can be used to compute  entanglement Renyi entropies:
\be
\label{renyicomp}
S_{n} = - \frac{1}{n-1} \left( \ln Z_\mathcal{M} (ds^2) - n  \ln Z_1 \right)
\ee
where $Z_1$ is the partition function of the theory on the flat $z$ plane without any 
branch points. 

The isometries of the surface \eqref{rs} include $\mathbb{Z}_n$ cyclic rotations
of the replicas and the anti-holomorphic involution which reflects
about the real $z$ axis (the symmetry associated to complex conjugation
due to the fact that the $z_i$ all lie on the real axis.)
Together these generate the dihedral group $D_n$ and we
refer to this as the ``replica symmetry''. For more discussions
on the relevance of these symmetries to computations of the ERE see 
\cite{Headrick:2012fk}. 

It is common to think of $Z_{\mathcal M}$ as the correlation function
of twist operators in the  product orbifold theory
of $n$ copies of the CFT under consideration:
\be
\label{twists}
Z_{\mathcal M} \propto \left< \sigma_1 (z_1)  \sigma_{-1} (z_2) 
\ldots \sigma_1 (z_{2N-1})  \sigma_{-1} (z_{2N}) \right> 
\ee
up to some regulator factors which deal with the divergences associated
to the conical singularities. The twist operator $\sigma_{1}$ enacts the generator of cyclic permutation
of the $n$ CFTs upon circling it.  And the operator $\sigma_{-1}$ acts inversely
to $\sigma_1$. See for example \cite{Lunin:2000yv} whose results are relevant for computations of EREs for general CFTs.  The dimension of these twist operators is fixed
by the central charge $c$ of the CFT:
\be
\label{dimtwist}
h_n = \frac{c n }{12} \left( 1 - \frac{1}{n^2} \right)
\ee

\section{Prescription}
\label{sec:pre}

In this section we give a prescription for finding and computing certain saddles of
3 dimensional Einstein gravity that contribute to $Z_\mathcal{M}$ by the usual rules of AdS/CFT. 
Many things will be introduced in an ad-hoc way leaving
their justification to later. We also leave discussions of the explicit bulk solution
to later sections. 

The prescription reproduces several known cases
as well as the Ryu-Takayanagi formula. Throughout this section
we will assume that one of the saddles we construct is the dominant solution,
and thus at leading order in $1/G_N$ computes the ERE. It should be kept in mind that this might not be the case. And so the prescription given here remains to be proven. 

We claim that in order to compute $S_n$ holographically one should use the following
recipe:
 
\begin{enumerate}
\item Consider the ordinary differential equation (ode) defined on the Euclidean
$z$-plane with the points $z_i$ lying on the real axis:
\be
\label{fuchs}
\psi''(z) + \frac{1}{2} T_{zz} \psi(z) = 0 \,; \qquad
T_{zz} = \sum_{i=1,\ldots 2N }\left(  \frac{\Delta}{(z-z_i)^2} + \frac{p_i}{z-z_i}\right)
\ee
where $\Delta =  ( n^2- 1)/(2n^2)$. The $p_i$ are called \emph{accessory} parameters. 

\item Tune $p_i$ such that the solutions of \eqref{fuchs} have \emph{trivial} monodromy around  a set of $N$ cycles $C_M$
in the z-plane with the points $z_i$ removed. We label this set by,
\be
\Gamma = \{ C_M : M = 1,\ldots N \}
\ee
The $C_M$ are defined to be simple non-intersecting (homologically) independent and
non-trivial and each encircle an even number of the $z_i$. At fixed $N$ there is some
number $\mathcal{N}_N$  of independent configurations of cycles $\Gamma_\gamma$ which we label by 
\be
 \mathcal{T}_N = \left\{ \Gamma_{\gamma} :\, \gamma = 1, \ldots \mathcal{N}_N \right\}
\ee

\item For a fixed configuration $\Gamma_\gamma \in \mathcal{T}_N$ the monodromy conditions determine the
$p_i^\gamma$. From these construct the following ``saddle'' Renyi entropies $S^\gamma_n$ by
integrating:
\be
\label{toint}
\frac{\partial S^{\gamma}_n}{\partial z_i} =  - \frac{c n}{6 (n-1)}  p_i^\gamma
\ee
where $c$ is the central charge.
\item The true ERE is claimed to satisfy:
\be
S_n = \min_{\gamma } S^\gamma_n
\ee

\end{enumerate}

We give some clarifying comments:
\begin{itemize}

\item The solution $\psi(z)$ will later be used to construct a bulk gravitational
solution. 

\item Prior to imposing the monodromy conditions the accessory parameters are real and unconstrained except for the three conditions:
\be
\label{acccond}
\sum_i p_i = 0\,, \qquad \sum_i p_i z_i = - 2 N \Delta\,,
\qquad \sum_i p_i z_i^2 = -2 \Delta \sum_i z_i
\ee
such that the point $z=\infty$ is not a singular point of the ode. Thus 
the point $z=\infty$ has trivial monodromy and one can think of \eqref{acccond} as being
contained within the monodromy conditions on the cycles in $\Gamma_\gamma$.

\item The counting of the number of unique configurations of cycles proceeds
recursively. As we add one more interval $N-1 \rightarrow N$ we can 
use configurations $\mathcal{T}_{N-1}$  
to construct those in $\mathcal{T}_{N}$. This is illustrated in Figure~\ref{countfig}.

\begin{figure}[h!]
\centering
\includegraphics[scale=.5]{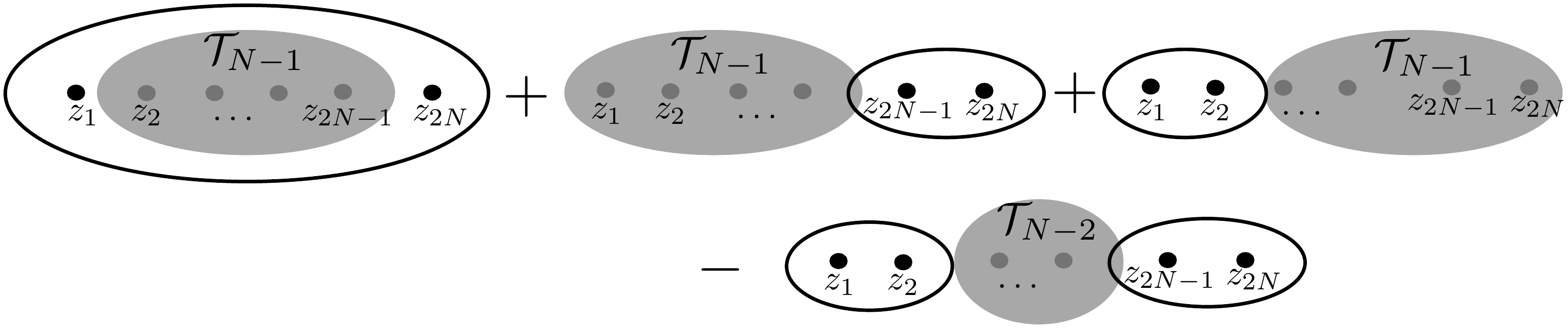}
\caption{A recursive argument for generating configurations
of cycles in $\mathcal{T}_N$. The black solid lines are new curves. The
other curves are represented by the shaded blob and are
taken from a configuration $ \Gamma \in \mathcal{T}_{N-1}$ or $\mathcal{T}_{N-2}$ as
indicated. The last term subtracts off some over counting of the previous two  terms.
The answer is $\mathcal{N}_N = 3 \mathcal{N}_{N-1} - \mathcal{N}_{N-2}$.
\label{countfig}}
\end{figure}

\item The condition that each cycle encircles an even number of points $z_i$ 
is related to the fact that these cycles actually live on the Riemann surface $\mathcal{M}$ and we want
them to come back to the same sheet.\footnote{Note the non-crossing condition
on the cycles is still appropriate despite the fact that some of the cycles actually
move into the second sheet. This follows from the  comment on cyclic symmetry.}

\item We will sometimes refer to a given $\gamma$ as a \emph{saddle} since it
will ultimately correspond to a particular three dimensional gravitational solution.
The monodromy conditions on $C_M \in \Gamma_\gamma$ 
will tell us which cycles of the manifold $\mathcal{M}$
are contractable inside the bulk three dimensional handlebody solution. 

\item For a manifold of genus $(N-1)(n-1)$ we should pick $(N-1)(n-1)$ non-intersecting cycles
(out of $2(N-1)(n-1)$) to be contractable in order to specify a unique handlebody. We will
sometimes refer to these as ``A-cycles.'' 
So far we have specified $N-1$ of these not including one of the cycles in $\Gamma_\gamma$ which is not independent due to the monodromy condition at infinity in the $z$ plane \eqref{acccond}.  
As we will see the remaining cycles are related to these by demanding
the bulk solution respects the replica symmetry. That is we are also implicitly picking a basis
of A-cycles:
\be
\{ g^m(C_M): m=0,\ldots n-1;\, M = 1, \ldots N-1 \}
\ee
where $g$ enacts the
cyclic replica symmetry and moves $C_M$ to the adjacent sheet
of the branched covering. Note that not all of these cycles are independent because  $\sum_{m=0}^{n-1} \left[ g^m(C_M)\right] = 0$. This gives the desired $(N-1)(n-1)$ counting.

\item  It is easy to see that the anti-holomorphic involution (symmetry under complex conjugation) is also preserved by this choice of cycles. 

\item Saddles we are missing include ones where the monodromy condition
on the Riemann surface $\mathcal{M}$ do not obey the replica symmetry. These cannot be constructed by the ode \eqref{fuchs}
which must be generalized in an appropriate way.

\item Note that up to some constants $T_{zz}$ will be the expectation value of the stress tensor for the associated saddle. It is then clear that $\Delta$ is related to the dimension of twist operators
\eqref{dimtwist}.
Furthermore \eqref{toint} follows from applying the conformal Ward identity 
to $T_{zz}$ and comparing to the conformal transformation of the twist operator
correlation function \eqref{twists} (albeit on a saddle by saddle basis.)
We will derive \eqref{toint} later using the bulk action for the constructed solutions.

\item The central charge is related as usual \cite{brownh} to the bulk Newton's constant
$c = 3 /(2 G_N)$. The prescription above is for large central charge, otherwise the different bulk
solutions will all contribute to \eqref{sadsum} including the ones we have not constructed.

\item Each saddle will have a counterpart set of geodesics which we can
identify with a locally minimal surface of the RT prescription. These geodesics can be constructed by noting that the configuration of cycles $\Gamma_\gamma$ partitions the $z_i$ into pairs: 
\be
\label{pgama}
P_\gamma = \{ (z_i, z_j)_K;\, K =1,\ldots, N   \}
\ee
such that $(z_i,z_j) \in P_\gamma$ are either both inside or both outside every cycle $C_M \in \Gamma_\gamma$. Joining these pairs by geodesics
gives the counterpart RT saddle.  The homology condition which is
part of the RT prescription \cite{Nishioka:2009un} is satisfied for these geodesics. See Figure~\ref{configtort} for an example of this.

\end{itemize}

\begin{figure}[h!]
\centering
\includegraphics[scale=.5]{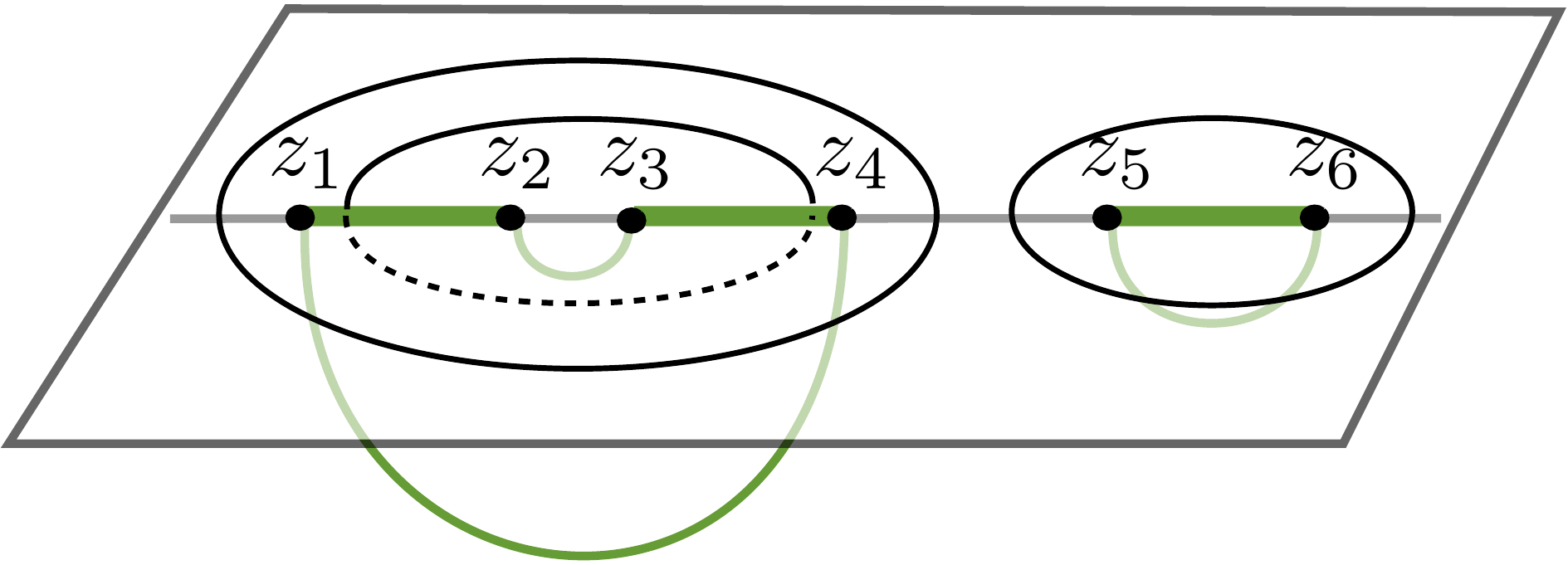}
\caption{A picture of the correspondence between boundary cycles for a fixed
configuration $\Gamma_\gamma \in \mathcal{T}_3$ and the bulk geodesics of the RT
formula (green curves hanging down from the boundary). The geodesics
connect points defined by $P_\gamma$ in \eqref{pgama}. Notice that in this picture the cycles
in $\Gamma_\gamma$ are contractable in the bulk without crossing the geodesics.
\label{configtort}}
\end{figure}

As a zeroth order check we consider $N=1$ where we find
that the conditions \eqref{acccond} are sufficient to fix the $p_i$,
\be
p_1 = - p_2 =   - \frac{2 \Delta}{(z_1-z_2)}
\ee
There is only a single configuration and it is clear that the monodromy is
trivially fixed to zero when passing around the points $z_1$ and $z_2$.
Integrating this we find the standard CFT result \cite{Holzhey:1994we}
\be
S_n\[N=1\] =  \frac{ c}{6} \left(1+\frac{1}{n}\right) \ln((z_2 -z_1)/\epsilon) + \kappa_{N=1}
\ee
where $\epsilon$ is a UV cutoff $\kappa_{N=1}$ is unfixed, but scheme dependent.

In general it is a difficult problem to carry out the above steps for $N>1$. Firstly there is no
analytic way to solve for $p_i^\gamma$ given a monodromy condition $\gamma$,
so one needs to proceed numerically. Secondly one needs to integrate \eqref{toint}
to find $S_{n}^\gamma$. We assume that the partial derivatives commute so
we only have a single integration constant for each saddle. Clearly the issue here is that the relative integration constants for the  saddle entropies are not fixed by the above prescription. 
We will fix these constants by taking limits where the saddle entropies $S_n^\gamma$ are related to the results for one less interval. This allows us to fix the integration constant recursively.

Note also that we can give an absolute formula for computing $S_n^\gamma$ which
will fix this integration constant, see \eqref{master} for the two interval case. The formula is written in terms of the solution $\psi$ to the ode problem but is more complicated than the prescription given above so we leave that till a later section. The prescription we have given is sufficient for
our current purposes.

\subsection{Reproducing the Ryu-Takayanagi prediction}

We wish to compute the monodromy matrices
in the replica limit $\delta_n  = n -1 \rightarrow 0$. We can do this
using perturbation theory.  Assume that
$p_i^\gamma$ vanishes linearly in the replica limit:
$ p_i^\gamma \sim \rho_i \delta_n   + \mathcal{O}(\delta_n^2)$
where $\rho_i$ are  constants which we need to determine.
The fact that $p_i^\gamma$ should vanish in the replica limit follows from
the conjectured formula for the entanglement Renyi entropy \eqref{toint} which we expect
to be finite in this limit. Also note that $\Delta  = \delta_n + \mathcal{O}(\delta_n^2)$ in this limit. The 
second order ode \eqref{fuchs} can be conveniently represented as a first order system:
\be
\frac{d}{d z} \begin{pmatrix} \psi \\ \psi' \end{pmatrix} = \begin{pmatrix}
0 & 1 \\ - \frac{1}{2} T_{zz} & 0 \end{pmatrix} \begin{pmatrix} \psi \\ \psi' \end{pmatrix}
\quad \rightarrow  \quad \frac{d}{dz} u(z) = H(z) u(z)
\ee
The monodromies are then simply path ordered exponentials:
\be
u(z) = M(C) u(z_0) \qquad M(C) = \mathcal{P} 
 \exp\left( \int_C dz H(z) \right)
\ee
where $C$ is a specific path from $z_0$ to $z$. Note that $M(C)$ has unit
determinant which follows from the Wronskian condition of two solutions to the ode.
Perturbatively we have
\be
H = H_0 + \delta_n H_1 = \begin{pmatrix} 0 & 1 \\ 0 & 0 \end{pmatrix} - \frac{ \delta_n }{2} T_1  \begin{pmatrix} 0 & 0 \\ 1 & 0 \end{pmatrix}\,, \qquad T_1  \equiv \sum_{i} \left( \frac{1}{(z-z_i)^2}
 + \frac{\rho_i}{(z-z_i)} \right)
\ee
So we can then use time dependent perturbation theory methods to solve this problem
where $z$ is thought of as ``time''. Firstly move to the interaction picture:
\be
M(C) = M_0(z) M_I(z)  \quad \rightarrow \quad \frac{d}{dz} M_I(z) = \delta_n( M_0^{-1}H_1M_0)(z) M_I(z)
\ee
where the zeroth order solution to the ode $\psi_0 = A z + B$ can be used to find
the zeroth order monodromy matrix:
\be
M_0(z) = \begin{pmatrix} 1 & (z-z_0) \\ 0 & 1 \end{pmatrix} 
\ee
The path ordered exponential expression for $M_I$ can then be computed to first order
by simply expanding the exponential:
\be
\label{monint}
M_I \approx \mathds{1} + \delta_n \int_C dz M_0^{-1}H_1M_0
 = \mathds{1}+\frac{\delta_n}{2} \int_C  dz \begin{pmatrix} (z-z_0) & (z-z_0)^2 \\ 1 & -(z-z_0) \end{pmatrix}  T_1
\ee
If we close the cycle $C$ by sending $z \rightarrow z_0$ we find the monodromy condition requires the vanishing
of resulting contour integral in Eq.~\ref{monint}.
For a cycle $C = C_M \in \Gamma_\gamma$ we get three independent conditions:
\be
\sum_{z_i \in D^\gamma_m} \rho_i = 0 \,,\quad
\sum_{z_i \in D^\gamma_m} (\rho_i z_i + 1 )  = 0 \,,\quad
\sum_{z_i \in D^\gamma_m} (\rho_i z_i^2 + 2 z_i )  = 0 \,,\quad
\ee
where the sum is over points $z_i$ contained in the interior of $C_M$
which we have denoted by the domain $D_M$. Note that it does not matter which ``interior'' we choose - because of the monodromy condition at $\infty$ given in
\eqref{acccond}. After some thought it becomes clear that this set of $N$ equations 
is solved  by the following conditions on the pairs $(z_i,z_j) \in P^\gamma$
into which the cycles $C_M$ partitioned the $z_i$.
\be
\rho_{l} = - \frac{2}{z_i -z_{j}} \,, \quad \rho_{j} = - \frac{2}{z_{j} - z_i}
\qquad \forall\,\, (z_i,z_j) \in P^\gamma
\ee
The saddle entanglement entropy can then be found by taking $\lim_{n \rightarrow 1}$
in \eqref{toint} and
integrating the result:
\be
S_{EE}^\gamma = \frac{c}{3} \sum_{(z_i,z_j) \in P_\gamma} \ln(|z_i-z_j|/\epsilon) + \kappa^\gamma_N
\ee
This result is exactly $c/6$ times the regulated lengths of geodesics in $AdS_3$ connecting
the points  $(z_i,z_j) \in P^\gamma$ on the boundary. As we discussed around
Figure~\ref{configtort} there is
a correspondence between the saddles that we construct (at any integer $n\geq 2$) and the
minimal surfaces (geodesics) needed to compute the RT answer. We have shown here
that the action of these saddles can be continued to non-integer $n$ and 
in the limit $n \rightarrow 1$ they become the lengths of the corresponding RT geodesics. 
While other aspects of this section are somewhat conjectural, the last statement
is correct and hints at the inner workings of the RT formula. 

To completely reproduce the RT prescription we are left to compute the integration constants $\kappa_\gamma$ relative to all the different
saddles $\Gamma_\gamma \in \mathcal{T}_N$. We give the following argument. Firstly consider an adjacent pair $(z_k,z_{k+1}) \in P_\gamma$ which is enclosed by a unique single cycle $C_L \in \Gamma_\gamma$ which does not enclose any other $z_i$
. Note that there must be at least one such
pair. Now take the limit $z_k \rightarrow z_{k+1}$ where we
expect to reproduce the entanglement entropy for $N-1$ intervals and 
a configuration of cycles given by $( \Gamma_{\gamma'} = \Gamma_\gamma \backslash C_L)
\in \mathcal{T}_{N-1}$. Atleast up to a UV divergence associated with the closing of the interval
$[z_k,z_{k+1}]$. For a very small interval $z_k \approx z_{k+1}$ we can zoom in on this and ignore all the other intervals
 - allowing us to exactly subtract off the EE associated
with this single interval. Note that it might be that $\[z_k,z_{k+1}\]$ is not
an interval in $\mathcal{A}$ but is an interval in the complement $\mathcal{A}^c$. In
which case 
we can appeal to approximate purity of the state at small distances so this still contributes the same divergence.
Further we require that we are in a regime where $S_{EE}^{\gamma'}$ is the dominant
saddle - this should be possible to arrange for by moving around the other $z_i$.
We find that
\be
\label{relt}
\lim_{z_k \rightarrow z_{k+1} } \left( S_{EE}^{\gamma} -  \frac{c}{3} \ln((z_{k+1} - z_k)/\epsilon) - \kappa_0 
\right) = S_{EE}^{\gamma'} 
\ee
Note we are assuming that the regulator we use is such that it treats the UV divergences
located at the different points $z_i$ in a uniform way. This way we get exactly $S_{EE}^{\gamma'}$
on the right hand side of \eqref{relt} and no other ambiguous constants. 
Assuming that $\kappa_{\gamma'} = (N-1)\kappa_1$ is fixed for
all configurations in $\mathcal{T}_{N-1}$ then by induction
we find $\kappa^{\gamma}_{N} = N \kappa_1$ which also must hold for all saddle configurations
in $\mathcal{T}_N$. The final answer: $\min_{\gamma} S^\gamma_{EE}$ is then the RT formula for disjoint intervals in a 2d CFT.

\subsection{Two intervals and the mutual information}

\begin{figure}[h!]
\centering
\includegraphics[scale=.6]{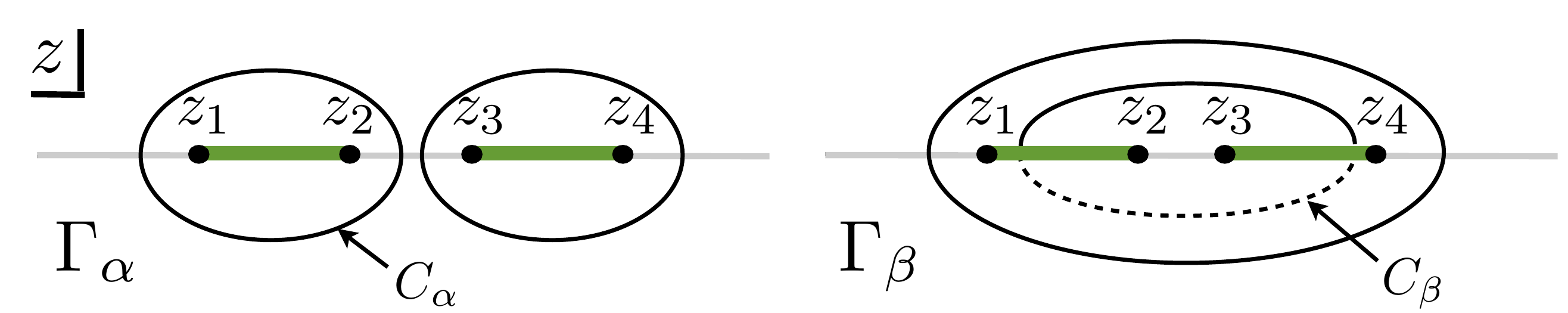}
\caption{
\label{fig2cyc} The case for two intervals. The black sold curves show the monodromy cycles
 $\Gamma \in \mathcal{T}_2$. There are two sets of cycles which we label $\alpha,\beta$.  The dashed curve on the right figure is due to the fact that this curve actually moves
through the branch cut into the last ($n$'th) replica. We also define
here the specific cycles $C_\alpha, C_\beta$ for later reference.} 
\end{figure}

We now specialize to the case of two intervals $N=2$. We think that
most of the following results work for $N>2$ but the arguments become cumbersome
and we content ourselves to looking in more detail at the first non-trivial case.
According to our prescription we have two different configurations
of cycles which we label $\gamma = \alpha,\beta$. These cycles 
are shown in Figure~\ref{fig2cyc}.
For example they correspond to the following partitioning of the $z_i$ into pairs:
\be
P_\alpha=\{(z_1,z_2),(z_3,z_4)\}\,,\quad P_\beta = \{(z_1,z_4),(z_2,z_3)\}
\ee
For ease of notation we will often drop the $\gamma=\alpha,\beta$ subscript when
the distinction is not important.

To start we would like to understand more about the dependence of the prescription
on the $z_i$. The conformal transformations
that leave the $z$ plane vacuum invariant
and also leave the $z_i$ on the real axis are given by
$SL(2,{\mathbb R})$ transformations:
\be
z \rightarrow \frac{ A z + B}{C z + D} \,, \quad z_i \rightarrow \frac{ A z_i + B}{C z_i + D}\,
\qquad AD -B C = 1 
\ee
where $A,B,C,D$ are all real. These move around the points $z_i$ and do not change the ordering up to cyclic permutations.  We would now like to track the transformation property of
$p_i$ in \eqref{fuchs} under $SL(2, {\mathbb R})$.
Firstly the conditions  \eqref{acccond} which the $p_i$ satisfy
are left invariant if we transform:
\be
\label{sl2t}
p_i \rightarrow (C z_i + D)^2 \left( p_i + 2\Delta \frac{ C}{C z_i + D} \right)
\ee
This property makes it clear that $p_i$ transforms
almost like a differential $\partial_{z_i}$. More precisely according
to \eqref{toint} $p_i$ is conjugate to $z_i$ and so in order to reproduce the transformation given in \eqref{sl2t} we
must demand that the entanglement entropy is not $SL(2, \mathbb{R})$ invariant
but rather:
\be
S_n \rightarrow S_n -  \frac{c }{6} \left(1 + \frac{1}{n} \right) \sum_{i=1}^4 \ln ( C z_i + D)
\ee
These statements follow trivially from the fact that ERE can be represented as
the $\log$ of a four point function of twist operators \eqref{twists}. Then the $SL(2, {\mathbb R})$
transformations above are simply due to the conformal weights of the 
twist operators. 

In order to soak up this transformation we can define what is known as
the Mutual Renyi Information (MRI). The MRI is the following combination
of entanglement Renyi entropies:
\be
I_n \equiv I_n([z_1,z_2] , [z_3,z_4]) = S_n([z_1,z_2]) +S_n([z_3,z_4])
- S_n([z_1,z_2] \cup [z_3,z_4])
\ee
such that:
\be
\label{midef}
I_n =  -  S_n  +  \frac{c}{6}  \left(1 + \frac{1}{n} \right) \ln\left( (z_2-z_1)(z_4-z_3)/\epsilon^2\right)
+ 2 \kappa_1
\ee

It follows that $I_n$ is $SL(2,\mathbb{R})$ invariant and as such
can only depend on the cross ratio:
\be
x = \frac{ (z_1-z_2)(z_3-z_4)}{(z_4-z_2)(z_3-z_1)}
\ee
From this we can define the following $SL(2,\mathbb{R})$ invariant accessory parameter:
\be
\label{toint2}
 \frac{d I_n(x)}{d x} = -   \frac{ c n}{6 (n-1)} p_x
\ee

There is one final property we have not exploited. Since we are working
in vacuum (to define our density matrix $\rho_\mathcal{A}$) it follows that the Renyi entropies
satisfy $S_\mathcal{A} = S_{\mathcal{A}^c}$ where $\mathcal{A}^c$ is the complement
of the region $\mathcal{A}$. This implies the following:
\be
S_n([z_1,z_2] \cup[z_3,z_4]) = S_n([z_4,z_1] \cup[z_2,z_3])
\ee
This purity relation corresponds to switching $z_4 \leftrightarrow z_2$ and as such
can be thought of as a very simple crossing relation for the twist operators.
When we  plug this into the mutual information
we find:
\be
I_n(x) = I_n(1-x) + \frac{c}{6}  \left(1 + \frac{1}{n} \right) \ln\left(\frac{x}{1-x}\right)
\ee

Actually we can go a little further. If we track what happens to the configurations 
of cycles in $\mathcal{T}_2$
as we send $z_4 \leftrightarrow z_2$ it is clear that $\Gamma_\alpha \leftrightarrow \Gamma_\beta$. So in
terms of the saddle ERE we can define the  saddle MRI
which satisfies:
\be
\label{purity}
I_n^\beta(x) = I_n^\alpha (1-x) + \frac{c}{6}  \left(1 + \frac{1}{n} \right) \ln\left(\frac{x}{1-x}\right)
\ee
and similarly for $\alpha \leftrightarrow \beta$.

In all we can now refine our prescription a little more. We can make
a conformal transformation to move the points to $z_1=0,z_2=x,z_3=1,z_4=\infty$.
After which our ode looks like:
\be
\psi''(z) + \frac{1}{2} \left( \frac{\Delta}{z^2} +\frac{\Delta}{(z-x)^2} + \frac{\Delta}{(z-1)^2} - \frac{2 \Delta}{z(z-x)}  - \frac{p_x x(x-1) }{z(z-1)(z-x)} \right) \psi(z) = 0
\ee 
and the prescription to compute the mutual information is simply
to integrate \eqref{toint2}. We should then find the maximum of the two
possible saddle mutual informations (note the sign switch in the definition \eqref{midef} of $I_n$):
\be
I_n(x) = \max\{I_n^\alpha(x), I_n^\beta(x) \}
\ee
where we remind the reader that it is possible there are some missing saddles which become dominant at some $x$ and thus override this answer.  

It is clear that in the  limit where $x\rightarrow 0$ the dominant configuration is $\Gamma_\alpha$ and in this limit the mutual information vanishes since this limit corresponds to moving the two intervals
infinitely far apart.  This condition will be used to fix the integration
constant in \eqref{toint2}.

Since when $x=1/2$ the two different saddle mutual informations
agree (by purity of the vacuum state) it must be the case that the $\Gamma_\alpha$ and
$\Gamma_\beta$ saddles switch dominances at $x=1/2$ \cite{Headrick:2010zt}. This
results in a first order phase transition for any $n$.

If we are feeling
lazy we can reconstruct the calculated contribution to the mutual information from a single saddle:
\be
I_n(x) = \left\{ 
\begin{aligned}
&I_n^\alpha(x) \,, &\qquad 0<x< 1/2 \\
&I_n^\alpha(1-x) + \frac{c}{6}  \left(1 + \frac{1}{n} \right) \ln\left(\frac{x}{1-x}\right) \,, &\qquad 1/2<x< 1
\end{aligned} \right.
\ee

\subsection{Reproducing the known answer for $n=2$} 

Set $n=2,\Delta = 3/8$ and it turns out in this limit 
we can analytically solve the ode. The reason lies in the fact that we are in this case secretly describing a genus one torus. The case for $n=2$ was  already worked out in \cite{Headrick:2010zt} based on fairly extensive computations given in \cite{Lunin:2000yv}. We will see that our prescription reproduces their results with relative ease.

The two indepdenent solutions can be written as:
\be
\psi(z) = \frac{1}{t'(z)^{1/2} }\exp\left(\pm h t(z)\right)\,,
\qquad t'(z) = \frac{1}{\sqrt{ z(z-1)(z-x)}}
\ee
where $h$ is an unfixed constant which is related to the accessory parameter:
\be
p_x = \frac{2-x}{4x(x-1)} + \frac{h^2}{2 x (x-1)}
\ee
These solutions can then be used to find the monodromy matrix:
\be
M(C) = \Psi(z) \begin{pmatrix} e^{ h \int_C t'(z) dz} & 0 \\
0 & e^{ - h \int_C t'(z) dz } \end{pmatrix} \Psi(z_0)^{-1}
\ee
where $C$ is a path from $z_0$ to $z$ and the non-path dependent factors are:
\be
\label{psiref}
 \Psi(z) = \frac{1}{(t')^{1/2}} \begin{pmatrix} -1 & -1 \\  \frac{1}{2} \frac{t''}{t'} - h t'
&   \frac{1}{2} \frac{t''}{t'} + h t' \end{pmatrix}
\ee
The trivial monodromy condition for the curve $C \in \Gamma_\alpha$ is then simply:
\be
 2 \pi i k = h \int_C t'(z) dz = 2 h \int_0^x \frac{dz}{\sqrt{ z(z-1)(z-x)}} 
  = 4 h K(x) 
\ee
where $k$ is an integer and $K(x)$ is the complete elliptic integral (defined by the integral
above.) The integer $k$ is unfixed so far. The case $k = 0$ does not work
since one has to be carefully when taking $h \rightarrow 0$ due to the degeneration of the matrix $\Psi(z)$ given in \eqref{psiref} in this limit.
For $|k| \geq 2$ the solutions one finds involve a multiply wound uniformization
coordinates (see Section~\ref{sec:schott}) and should not
be included in our prescription since they will not correspond to sensible bulk solutions.
We are left with $k= \pm 1$ of which either gives the same answer.  The accessory parameter is:
\be
p^\alpha_x =   \frac{2-x}{4x(1-x)} - \frac{ \pi^2}{8  x (1-x) K(x)^2}
\ee
Which integrates to:
\be
I_2^\alpha = - \frac{c}{12} \log\left( 2^8 (1-x)/x^2 \right) + \frac{c \pi}{6} \tau_2
\ee
where we have defined the (purely imaginary) modular parameter for the underlying
torus:
\be
\tau_2 = \frac{K(1-x)}{K(x)}
\ee
and we have added an integration constant such that $I^\alpha_{2}(x=0)=0$.
Similarly we can find $I_2^\beta$ by imposing the different monodromy condition
on the cycles $\Gamma_\beta$. The answer one finds satisfies the expected
purity relation \eqref{purity} where under $x \rightarrow 1 -x$ the modular
parameter of the torus undergoes the modular $S$ transformation $\tau_2 \rightarrow 1/\tau_2$.
This makes sense because $S$ switches
the cycles on the torus and thus the two monodromy conditions we are working with.

\subsection{Numerics for $n>2$}

In order to compute the monodromy matrices numerically it is convenient to define
connection matrices along the real line between the singular points. These
matrices relate  canonically chosen linearly independent solutions at adjacent singular point.
We relegate the details to Appendix \ref{app:conn}. The monodromy condition can easily
be read off from these connection matrices and from this we can compute
$p_x$. 

\begin{figure}[h!]
\centering
\includegraphics[scale=.47]{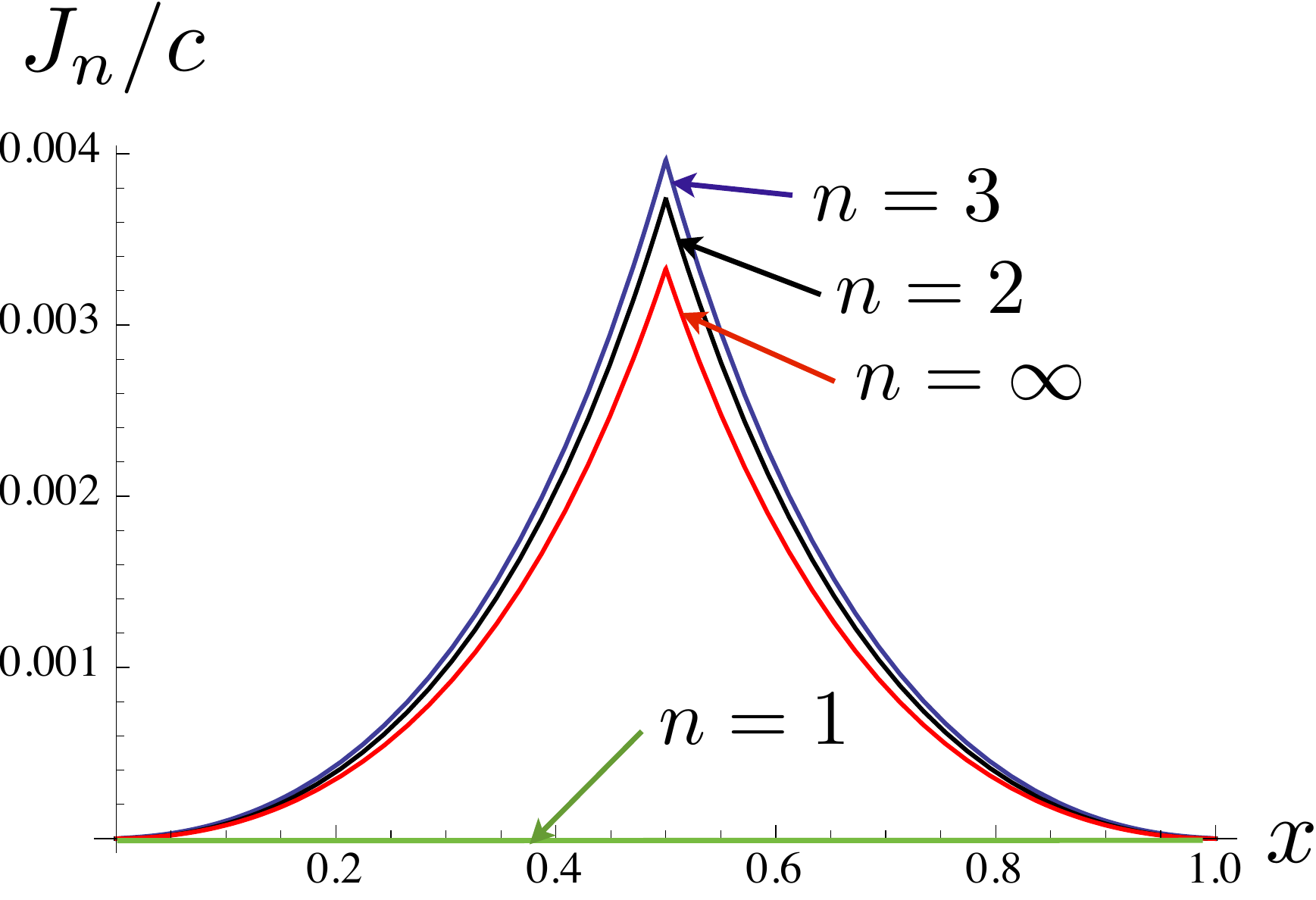} \includegraphics[scale=.44]{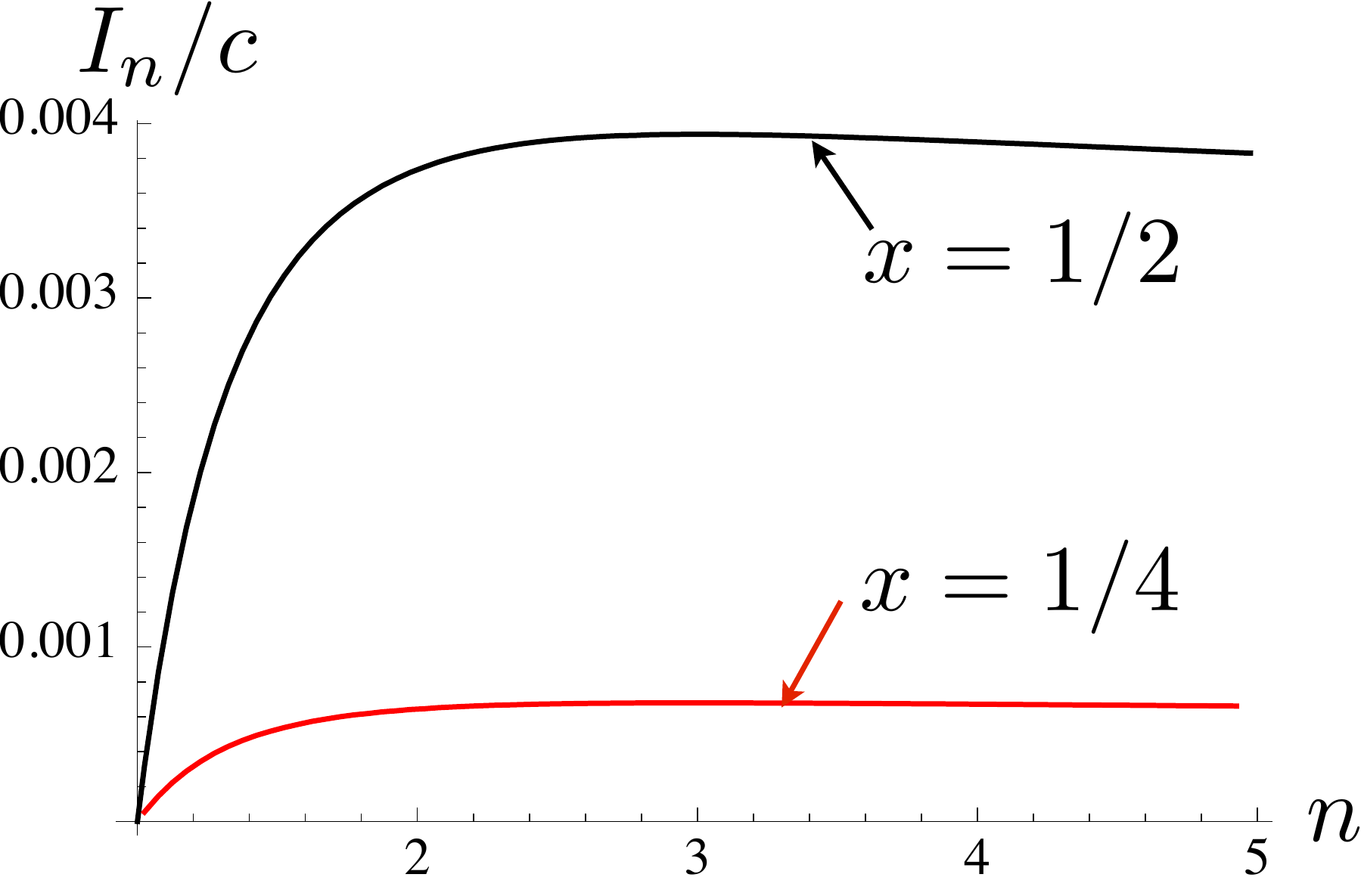}
\caption{Calculated contributions to the Mutual Renyi Information (MRI) in holographic CFTs.
We show in the \emph{left} panel a subtracted version of the MRI as defined in \eqref{subtracted}.
Both saddles $\Gamma_\alpha$ and $\Gamma_\beta$ are important and 
dominate for $x<1/2$ and $x>1/2$ respectively. In the \emph{right} we show
the dependence of the MRI on $n$ for fixed $x$ (after analytically continuing from
integer $n$). The $n=1$ limit for $x \leq 1/2$ 
is zero as predicted by the RT formula. Numerically it was more convenient to use
\eqref{master} to find this \emph{right} plot.
\label{fig:eres}
}
\end{figure}

The results are shown in Figure~\ref{fig:eres}. Actually there is very little difference
between the Mutual Renyi Information for different values of $n$ and in order to effectively 
compare them we subtract off a scaled version of the EE ($n=1$) which takes
into account the scaling of the twist operators with $n$:
\be
\label{subtracted}
J_n(x) = I_n(x) - \frac{1}{2} \left(1+\frac{1}{n} \right) I_1(x)
\ee
Recall that the RT formula for the MI is  $I_1(x) =  \max\{ 0,(c/3) \ln \frac{x}{1-x} \}$.
The function $J_n(x)$ has the property that it is symmetric about $x=1/2$.

\section{Schottky Uniformization}

\label{sec:schott}

Having introduced the ode \eqref{fuchs} as the main crux for
constructing certian bulk solutions we should now explain
where this came from, and in particular give some pictures of what the
bulk solution looks like.  Since all solutions of Einsteins equations with a negative cosmological constant in 3 dimensions  are globally quotients of $AdS_3$ we simply need to determine
the appropriate quotient. The technology we need in order to do
this goes under the name of Schottky uniformization. We give here a rough
general discussion of this technology. We follow closely the discussion in \cite{Krasnov:2000zq} and \cite{zta} see also \cite{maskit}.

Pick the following coordinates on $AdS_3$:
\be
\label{ads3}
ds^2 = \frac{ d\xi^2 + d w d \bar{w} }{\xi^2}
\ee
with conformal boundary at $ \xi \rightarrow 0$ the complex $w$ plane.
The isometry group of $AdS_3$ is $PSL(2,\mathbb{C})$ where the action induces
a conformal isometry on  the boundary. The action on $AdS_3$ is:
\begin{align}
\label{actionads3}
w \rightarrow \frac{ (a w + b) (\bar{c} \bar{w} + \bar{d} ) + a \bar{c} \xi^2}{ |c w + d|^2 + |c|^2 
\xi^2} \qquad   \xi \rightarrow \frac{\xi}{  |c w + d|^2 + |c|^2  \xi^2}
\end{align}
where $a d - bc = 1$. As $\xi \rightarrow 0$ this action becomes:
\be
\label{actionbdry}
w \rightarrow \frac{ a w + b}{ c w + d} \equiv L(w) \qquad \xi \rightarrow \xi |L'(w)|
\ee

In this way the quotient of $AdS_3$ by a discrete subgroup $\Sigma \subset PSL(2,\mathbb{C})$
descends to a quotient of the complex $w$ plane.\footnote{We are being heuristic here -
for example we should first remove a certain set of measure zero from the $w$ plane, for which $\Gamma$
acts badly (fixed points of $\Gamma$): $\mathbb{C}' = \mathbb{C}/\{\rm bad\,points\}$.
We can then form the quotient $\mathbb{C}'/\Sigma$.  For a proper discussion see. We will continue to be heuristic without making similar admissions. } 
This quotient is then a way of representating the surface
$\mathcal{M} = \mathbb{C}'/\Sigma$. Thus one of the steps we will need to understand is 
how to map the $w$ complex plane into $\mathcal{M}$: 
\be
\pi_S : \mathbb{C} \rightarrow \mathcal{M} 
\ee
consistent with the action of the quotient. In fact the ode \eqref{fuchs} is exactly what determines this map. At the same time the monodromies of the ode \eqref{fuchs} determine the correct quotients of the $w$-plane. 

Let's see roughly how this works.  Consider instead the inverse map $w = \pi_S^{-1}(z)$ which is  multivalued on $\mathcal{M}$. The Schwarizian derivative
of this  map behaves like a holomorphic CFT stress tensor on $\mathcal{M}$:
\be
\label{sch2}
 \left\{ \pi_S^{-1},z \right\}= \frac{w'''}{w'} - \frac{3}{2} \left(\frac{ w''}{w'} \right)^2 \equiv
 T_{zz}(z)
 \ee
This equation can be thought of as a differential equation for $\pi_S^{-1}(z) = w(z)$ where
$T_{zz}$ is taken as a fixed input. We will construct $T_{zz}$ independently in a moment. 
Solving  equation \eqref{sch2} is equivalent to solving a second order ode:
\be
\label{ratsoln}
w(z) = \frac{\psi_1(z)}{\psi_2(z)} \,, \qquad \psi'' +\frac{1}{2} T_{zz} \psi =0
\ee
where $\psi = \psi_{1,2}$ are two linearly independent solutions
of this ode.
At this point we have made a connection with the prescription
of Section~\ref{sec:pre}. However we still need
to give an argument that the tress tensor $T_{zz}$ takes the form quoted in \eqref{fuchs}.
Since the map $\pi_S^{-1}$ is globally defined (but multivalued) and
since the the Schwarzian derivative does not change under the  $PSL(2,\mathbb{C})$
action on $w$ the stress tensor is globally defined on $\mathcal{M}$ and not
multi-valued.  However $T_{zz}$ does not transform homogeneously under general 
conformal transformations $z \rightarrow z(\widetilde{z})$
since the Schwarzian derivative shifts under such coordinate changes:
 $\{t,\widetilde{z}\} = z'(\widetilde{z})^2 \{t,z\} + \{z,\widetilde{z}\}$. 
The  rule on the overlapping patches is:
\footnote{ Note that this also implies that the solutions of the ode transform as
$-1/2$ differentials:
\be
\widetilde{\psi}(\widetilde{z}) = \psi(z(\widetilde{z})) \left( \frac{ \partial z}{\partial \widetilde{z}} \right)^{-1/2}
\ee }
\be
\label{xform}
\widetilde{T}_{\widetilde{z}\widetilde{z}} = \left( \frac{ \partial z}{\partial \widetilde{z}} \right)^2 T_{zz}
+ \{ z,\widetilde{z} \}
\ee 

Given these properties we claim that the following expression (on the $z$ coordinate patch) is smooth
on $\mathcal{M}$ and completely general:
\be
\label{stress}
T_{zz} = \Delta \left( \sum_{i=1}^4 \frac{1}{(z-z_i)^2} 
+ \frac{2(-z_3+z_1+z_2+z_4-2 z)}{(z-z_1)(z-z_2)(z-z_4)} \right) + \sum_{s=1}^{3(n-2)} \hat{p}_s \omega^{s}_{zz}
\ee
where $\Delta = 1/2 (1-1/n^2)$. To argue for this form consider to begin with the
the last sum over $\omega^s_{zz}$  which are holomorphic 
quadratic differentials on $\mathcal{M}$.
There are $3(n-2)$ of these and they are enumerated in Appendix~\ref{app:rsurf}.
Given a fixed $T_{zz}$  we can always add a linear combination of quadratic differential since they
transform homogeneously under conformal transformations thus
preserving \eqref{xform}.  

Finally we have to check that the remaining term multiplying $\Delta$ in \eqref{stress}
 is smooth on $\mathcal{M}$. This term is \emph{not}
a quadratic differential  which are smooth  by definition.  
We only need to check the behavior near $z \rightarrow z_i$ and $z \rightarrow \infty$. 
Close to for example $z \rightarrow z_1$ we can use the coordinate
$y$ defined by the branched covering
\eqref{branch2} $y^n \sim (z-z_1)$ such that the Schwarzian derivative is:
\be
\{z,y\} \approx -\frac{1}{2} \frac{n^2-1}{y^2} + \mathcal{O}( y^{n-2} )
\ee
and thus the new stress tensor in the $y$ coordinate patch is:
\be
\widetilde{T}_{yy}(y) = \frac{1}{y^2} \left( \Delta n^2  -\frac{1}{2} (n^2-1) \right) +
\mathcal{O}( y^{n-2} )
\ee
This is smooth provided $\Delta = 1/2 (1-1/n^2)$ and $n \geq 2$. 
The last term in the brackets of \eqref{stress} is then required for
smoothness as $z \rightarrow \infty$.

The logic of the preceding discussions is that  we have replaced the problem of finding the map $\pi_S^{-1}$ with the problem of finding the accessory parameters $\hat{p}_s$.
As was discussed extensively in Section~\ref{sec:pre} these should be determined by the monodromy
conditions imposed on solutions to the ode.  For example if we traverse a closed path $C$ on $\mathcal{M}$ the map $w=\pi_S^{-1}$ defined by \eqref{ratsoln} is not single valued
since the solutions $(\psi_1,\psi_2)$ undergo a monodromy $M(C)$:
\be
(\psi_1,\psi_2) \rightarrow  (\psi_1,\psi_2) M(C) \quad \implies \quad
\omega\, \mathop{\rightarrow} \, \frac{ a \omega  + b}{ c \omega + d} \,, \qquad \begin{pmatrix} a & c 
\\ b & d \end{pmatrix} \equiv M(C)
\ee 

Thus the monodromies of the ode determine a $PSL(2,\mathbb{C})$ action
on $w$. And in this way they determine the discrete quotient group $\Sigma$.
Note that the monodromies form a representation of the fundamental group of the surface $\mathcal{M}$
and so does the quotient group $\Sigma$. 

We need to understand which groups $\Sigma$ produce
the desired handlebody solutions when acting on $AdS_3$. These groups are called Schottky groups and we simply quote some results. They
have the property that $\Sigma$ is freely generated by half of the $2(n-1)$ generators
in the fundamental group. We define these generators through their $PSL(2,\mathbb{C})$
representative: $\{ L_m: \, m= 1 \ldots n-1\}$. 
Upon traversing around the \emph{other} half of the generators
of the fundamental group one finds trivial monodromy and trivial action in the quotient group. These generators
correspond to a basis of non-intersection cycles in the homology of $\mathcal{M}$, 
a basis of ``A cycles''.   It turns out that these are the cycles which are contractable in the bulk of the corresponding $AdS_3$ handlebody. 
The ``B cycles'' then correspond to the generators
which have nontrivial action $L_m$ and are not contractable.

\begin{figure}[h!]
\hspace{-1.1cm}
\includegraphics[scale=.4]{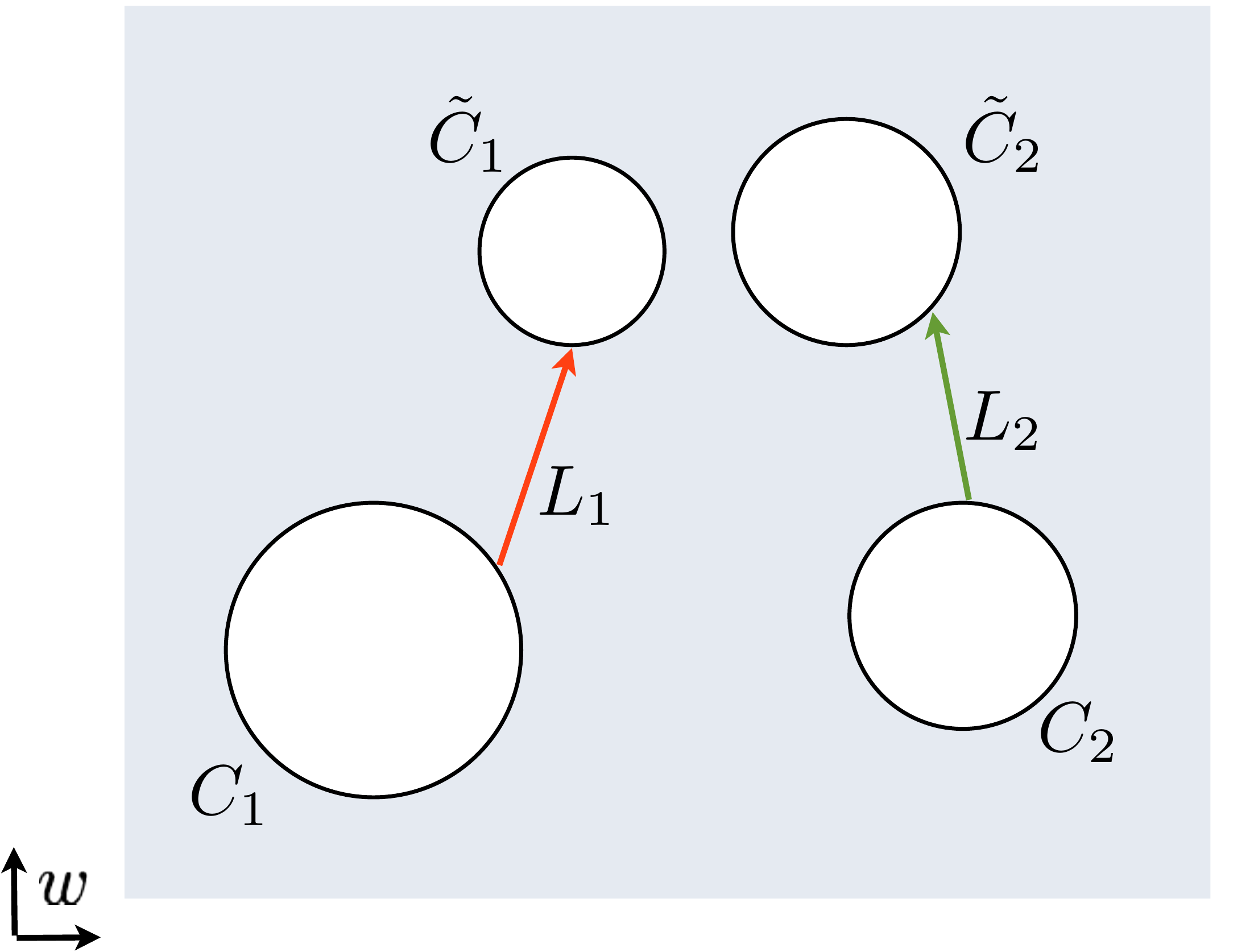}
\caption{
\label{sch-cartoon} 
The Schottky fundamental domain for a genus $2$ surface
with $2$ generators $L_1,L_2$. The shaded region is the domain
(continued to infinity.) Note that typically one normalizes the generators
using the freedom to conjugate by $PSL(2,\mathbb{C})$ such that
one of the circles surrounds $w = \infty$ which is then
absent from the domain. In the above picture we
have not done this, since this will be convenient for us later.
}
\end{figure}

We can get a rough picture of the quotient looking more carefully at
the fundamental domain of $\mathbb{C}'/\Sigma$ and $AdS_3'/\Sigma$.
See Fig.~\ref{sch-cartoon} for a picture. The fundamental domain  is given by specifying $(n-1)$ pairs of 
non-intersecting circles in the $w$ plane: $\{ C_m, \widetilde{C}_m : m=1 \ldots n-1 \}$ and then 
identifying the circles $C_m$ and $\widetilde{C}_m$ via the non-trivial generators $L_m(C_m) = \widetilde{C}_m$.
Note that the fundamental domain is not unique. The generators $L_m$ map the outside
of the circle $C_m$ into the inside of the circle $\widetilde{C}_m$. So for example 
we can shrink $C_m$ while making $\widetilde{C}_m$ larger and still have a fundamental domain for
the quotient.

The fundamental domain of the quotient of $AdS_3$ is simply found by extending
the circles $C_m, \widetilde{C}_m$ living on the boundary to hemispheres in the bulk
of \eqref{ads3}. 
These are two dimensional minimal surfaces in $AdS_3$ ending on the circles.
Note that on $\mathcal{M}$ the cycles which encircle $C_m, \widetilde{C}_m$
are contractable within the three dimensional bulk solution.  
These correspond to the cycles of $\mathcal{M}$ with trivial monodromies (the A-cycles.) 
The B-cycles are paths in the fundamental domain which connect the
identified circles.

Finally we come to the accessory parameters. In \eqref{stress} 
we have $3(n-2)$ of these, however we claimed in Section~\ref{sec:pre} that
there was only a single independent accessory parameter for two intervals $p_x$.
Note that of all $3(n-2)$ quadratic differentials enumerated in Appendix~\ref{app:rsurf}
only one of them $\omega_{zz}^1$ does not change under the actions of the replica symmetry.
This means that in order to preserve this symmetry $\hat{p}_s = 0 $ for $s \neq 1$. If we include the anti-holomorphic involution (complex conjugation on the $z$-plane) in the replica symmetry we find that the remaining accessory parameter $\hat{p}_1\equiv p_x$ should be real.  
Fixing most of the accessory parameters to zero is only possible if the monodromy
conditions respect the replica symmetry otherwise these should be turned on
and the resulting bulk solution will also not by symmetric.

\section{Bulk Solution}

\label{sec:bs}

We turn now to a detailed description of the bulk solution, from
which the final goal is to compute the bulk action which we get
to in the next section. We start with the details of the quotient $\mathbb{C}'/\Sigma$.

Assume the Schottky monodromy problem has been solved.
As discussed in the previous sections for two intervals  and for bulk solutions which are replica symmetric there are  two different monodromy conditions
that we can impose that we labelled $\Gamma_\alpha,\Gamma_\beta \in \mathcal{T}_2$
(see Figure~\ref{fig2cyc}). For arguments sake pick $\Gamma_\beta $ which involves imposing
trivial monodromy around a cycle $C_\beta$ which encircles the pair $(z_2,z_3)$. The other case can be worked out in an analogous manner. We would like to work out the identification circles $C_m, \widetilde{C}_m$ for the Schottky fundamental domain as well
as the generators $L_m$ linking them. The Schottky group $\Sigma$ is only
defined up to common conjugation by $PSL(2,\mathbb{C})$ and thus we
can choose two independent solutions of the ode at will in order to produce
the map $w(z)$. We pick the solutions,
\be
\psi_{\pm} = (z-z_1)^{1/2\pm1/2/n} (1 + \mathcal{O}(z-z_1) ) 
\ee
which diagonalize the monodromy around the point $z_1$. Then define:
\be
\label{sch}
w = \lambda \frac{\psi_+}{\psi_-}
\ee
such that $w(z_1)=0$. Note that under this choice the $\mathbb{Z}_n$ replica
symmetry is generated by rotations of the $w$ plane by an angle $2\pi/n$.
A nice way to get a concrete picture 
of the map generated by \eqref{sch}
is to consider the images under $w(z)$ of the $4$  
real axis segments in the $z$-plane between the points $z_i$.
Segments slightly above and slightly below the real $z$-axis $\pm i \eta$ 
should both be considered since
these will map to different curves in the $w$ plane. We should
also consider the segments on all the $n$ replicas. The images of these segments will
then trace out a particular fundamental domain in $w$.
The identifications $L_m$ can be worked out by appropriately glueing the
real line segments together amongst the different replicas.

\begin{figure}[h!]
\hspace{-.2cm} \includegraphics[scale=.44]{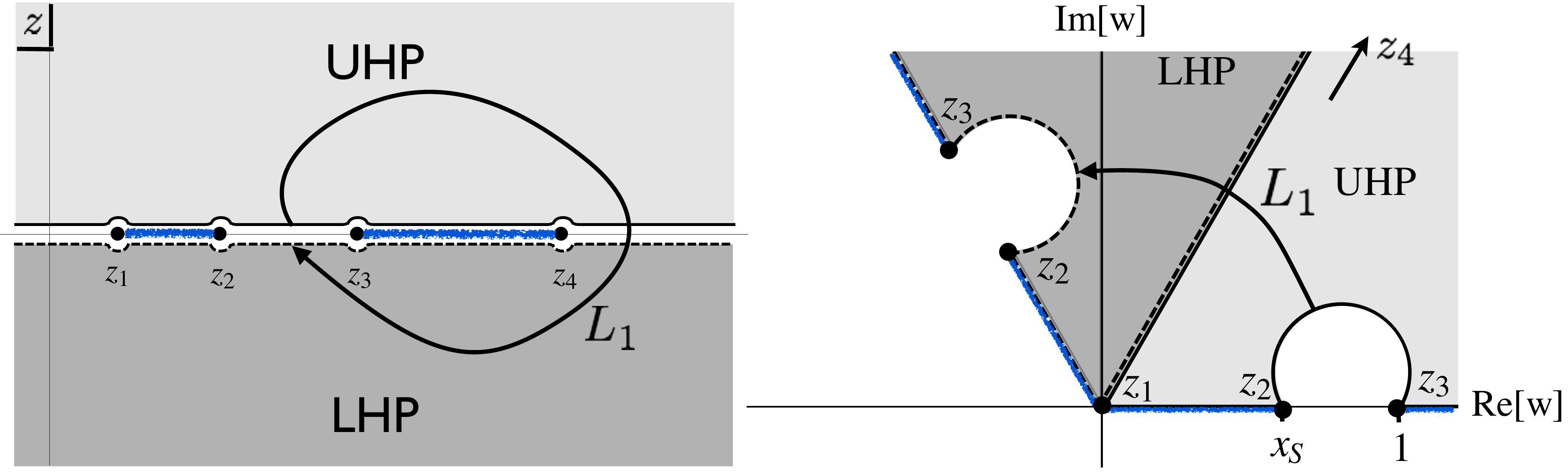}
\caption{A picture of the map from the branched covering $z$ (\emph{left}) to the
Schottky domain $w$ (\emph{right}) for $n=3$. We only show the image of the first replica.
The solid line tracks the real line segments of the $z$ plane which are slightly above the real axis
and the dashed line tracks the segments just below the real axis. The blue
fuzzy lines represent the branch cuts. The point $z_4$ maps to $w(z_4) = \infty$.
Note that we have imposed
the monodromy condition on the cycles in  $\Gamma_\beta$ (Figure~\ref{fig2cyc}) so that the map $w(z)$ jumps
discontinuously across the segment $[z_2,z_3]$ via the generator $L_1$.
\label{fig:zw}
}
\end{figure}

\begin{figure}[h!]
\begin{center}
\includegraphics[scale=.6]{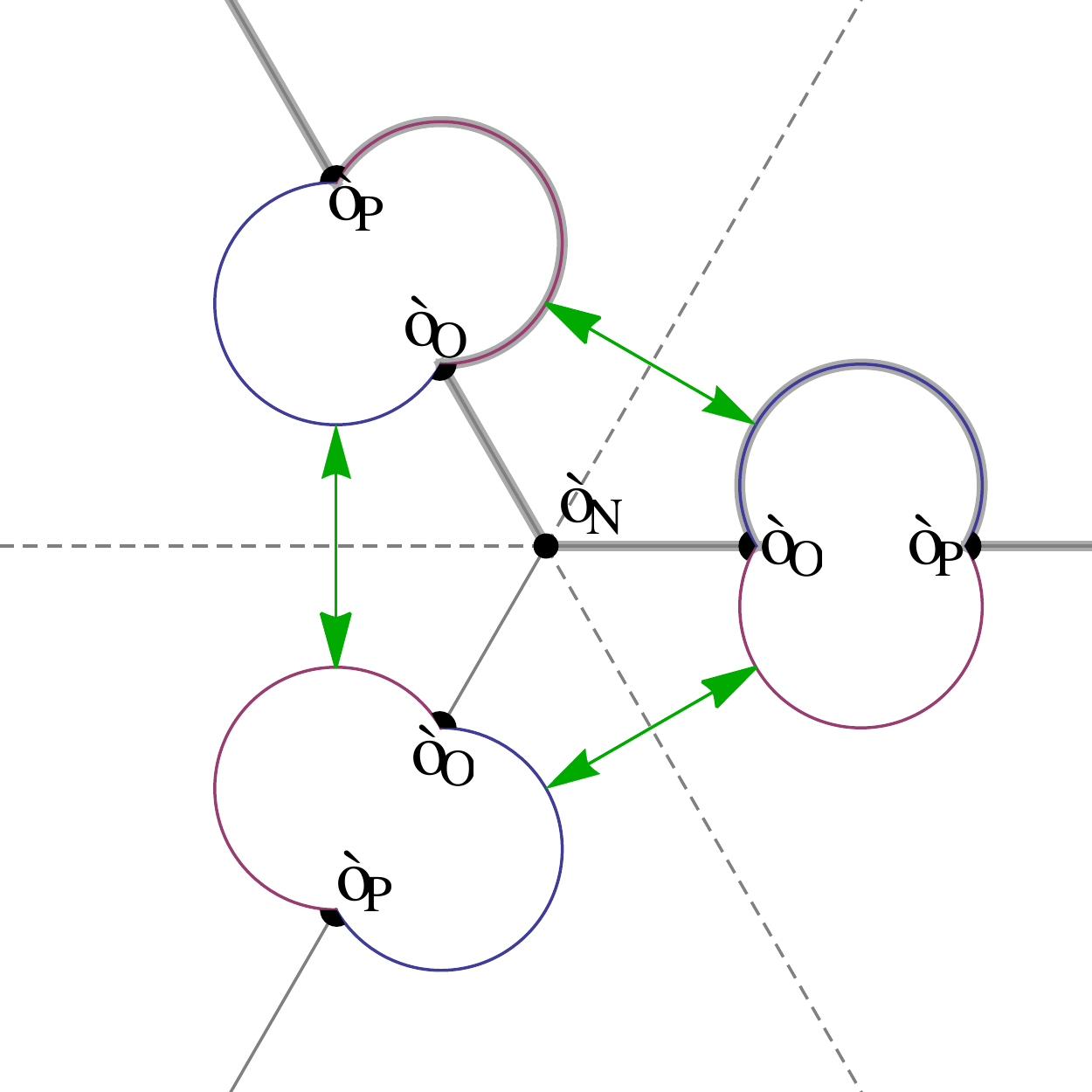} \includegraphics[scale=.6]{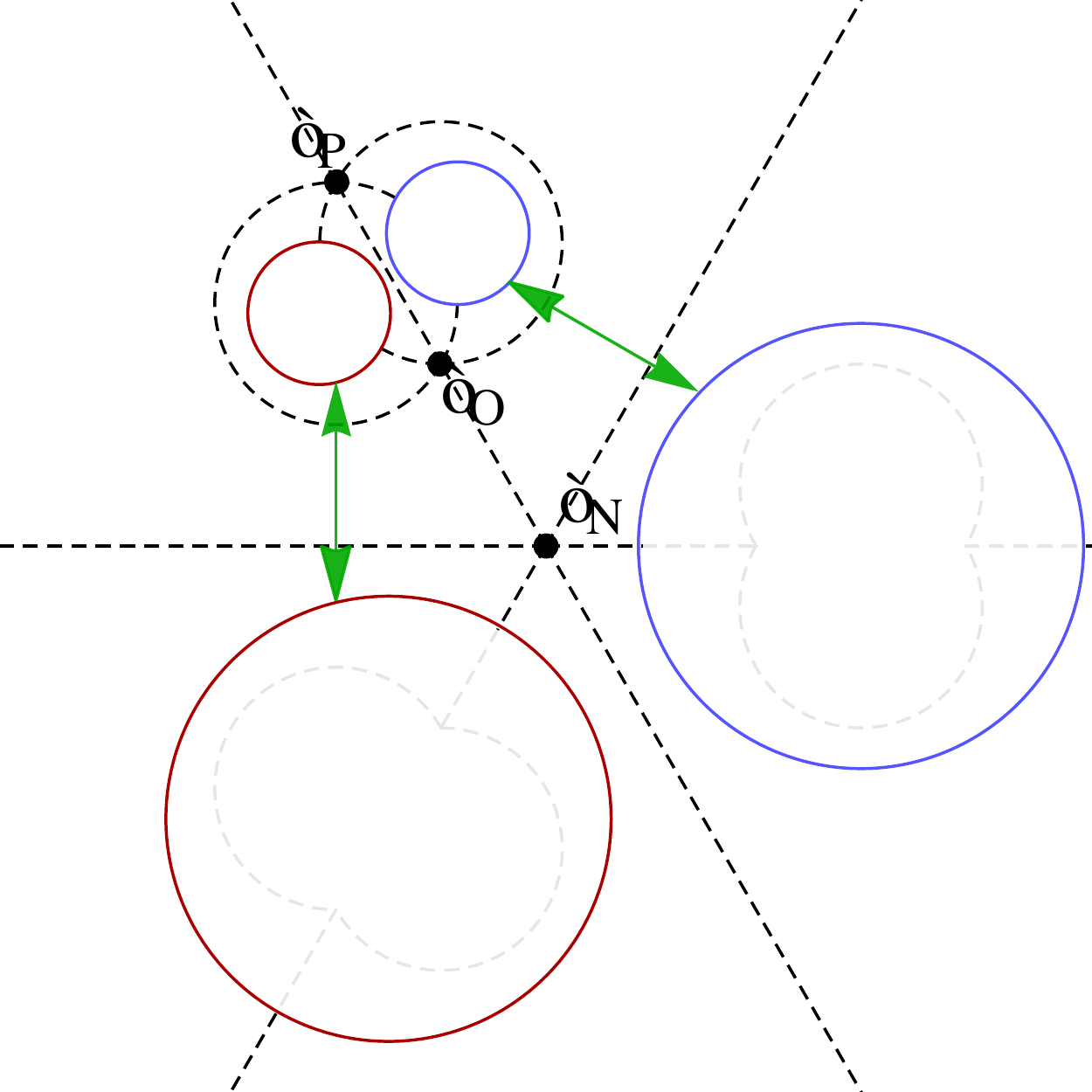}
\caption{Pictures of the fundamental domain of the Schottky quotient. We have
drawn the case $n=3$ for some fixed $x_S(x)$.
The \emph{left} plot shows a non-standard domain $\mathcal{D}_s$ which is however clearly
$\mathbb{Z}_n$ symmetric. There are three (generally $n$) generators $L_1,L_2,L_3$
for this case which are shown as green arrows. These generators are not all independent. 
In the \emph{right} picture we have deformed some of the circles in the \emph{left} picture to 
form a more standard fundamental domain $\mathcal{D}_d$  where there are now only
two generators $L_1, L_2$. Note that the images of the points $w(z_i)$ only appear once within
this fundamental domain. In the \emph{right} figure we label the blue circles $O_1, \tilde{O}_1$ and the red ones $O_2, \tilde{O}_2$ .
\label{fig:def}
}
\end{center}
\end{figure}

See Figure~\ref{fig:zw} for the resulting picture. This can be confirmed numerically
by plotting \eqref{sch} \emph{after} one has imposed the monodromy condition.
We will not give the full detailed argument that leads to this picture. However
we summarize some of the more important aspects:
\begin{itemize}
\item Since the ode \eqref{fuchs} is real along the real $z$-axis all the segment images
must be circular arcs or straight lines. That is there is always a basis of solutions
to the ode $\psi_I, \psi_{II}$ which is real along a given segment. Then:
\be
w_{\rm seg} = \frac{ a \ell + b}{ c \ell + d} \qquad \ell = \frac{\psi_I(z)}{\psi_{II}(z)} \in \mathbb{R}
\ee
for some $a,b,c,d \in \mathbb{C}$. This describes a circle or line in the
complex $w$-plane with affine parameter $\ell$.

\item Most of the real axis segments map to straight lines along rays emanating from
the origin in the $w$-plane $w = \ell \exp( i 2m \pi/n)$ and  $ w = \ell \exp( i (2m+1) \pi/n)$
where $m=1,\ldots n$ . In particular $w(z_4) = \infty$. This
behavior for \eqref{sch} only follows once the monodromy condition is imposed. 

\item Images of the segments $z\in [z_2,z_3] \pm i \eta$ on all the different replicas map to circular arcs which meet the above straight lines at an angle of $\pi/n$. These arcs can be
described as:
\begin{align}
\label{Um}
 w (\sigma+i\eta) &=  e^{ i 2\pi \frac{ (m-1)}{n}} \left( \frac{ x_S-  \ell  e^{-\frac{i \pi}{n}} }{1- \ell e^{-\frac{i \pi}{n}} } \right)  \qquad 0 < \ell  < \infty
\qquad z_2< \sigma <z_3 \\ 
\label{Utm}
 w(\sigma-i\eta) &=  e^{i2\pi\frac{ m}{n}} \left( \frac{ x_S-  \ell e^{\frac{i \pi}{n}} }{1- \ell e^{\frac{i \pi}{n}} } \right)  \qquad 0 < \ell < \infty \qquad z_2 < \sigma < z_3
 \end{align}
 where $m$ labels the images generated from the different replicas $m=1, \ldots n$.
 
\item We have chosen the magnitude of $\lambda$ in \eqref{sch} so that
$|w(z_3)|=1$.  Then $|w(z_2)| = x_S$ is the single remaining
parameter which can be computed numerical in terms of the cross ratio $x$.
It satisfies $ x_S <  1$.

\item Note that \eqref{sch} jumps discontinuously across  $[z_2,z_3]$ on the $z$-plane
since the arcs \eqref{Um} are different from \eqref{Utm}. This jump is encoding
a non-trivial monodromy element and occurs on one of the B-cycles.
Note in particular this jump does \emph{not}
occur across a branch cut on the $z$-plane. 

\item We can compute the associated
generator in $\Sigma$ 
by finding the $PSL(2,\mathbb{C})$ transformation which identifies an arc in  \eqref{Um} with the corresponding arc in  \eqref{Utm}
(with the same $m$.) By symmetry the two affine parameters $\ell$ map onto to each other. One finds:
\be
\label{gens}
L_m = \frac{1}{1-x_S} \begin{pmatrix} x_S   - e^{i \frac{2 \pi}{n}} & 2 i x_S e^{ i \pi \frac{ (2 m-1)}{n}}
 \sin(\pi/n)
\\  -2 i  e^{- i\pi \frac{(2m-1)}{n}}  \sin(\pi/n) &  x_S - e^{-i \frac{2 \pi}{n}}  \end{pmatrix}
\ee

\item Gluing the replicas together we get the global picture on the left
side of Figure~\ref{fig:def}.  Note that the circular arcs are the boundaries of the fundamental domain and they are identified pairwise as in the Figure.

\end{itemize}

\begin{figure}[h!]
\begin{center}
\includegraphics[scale=.55]{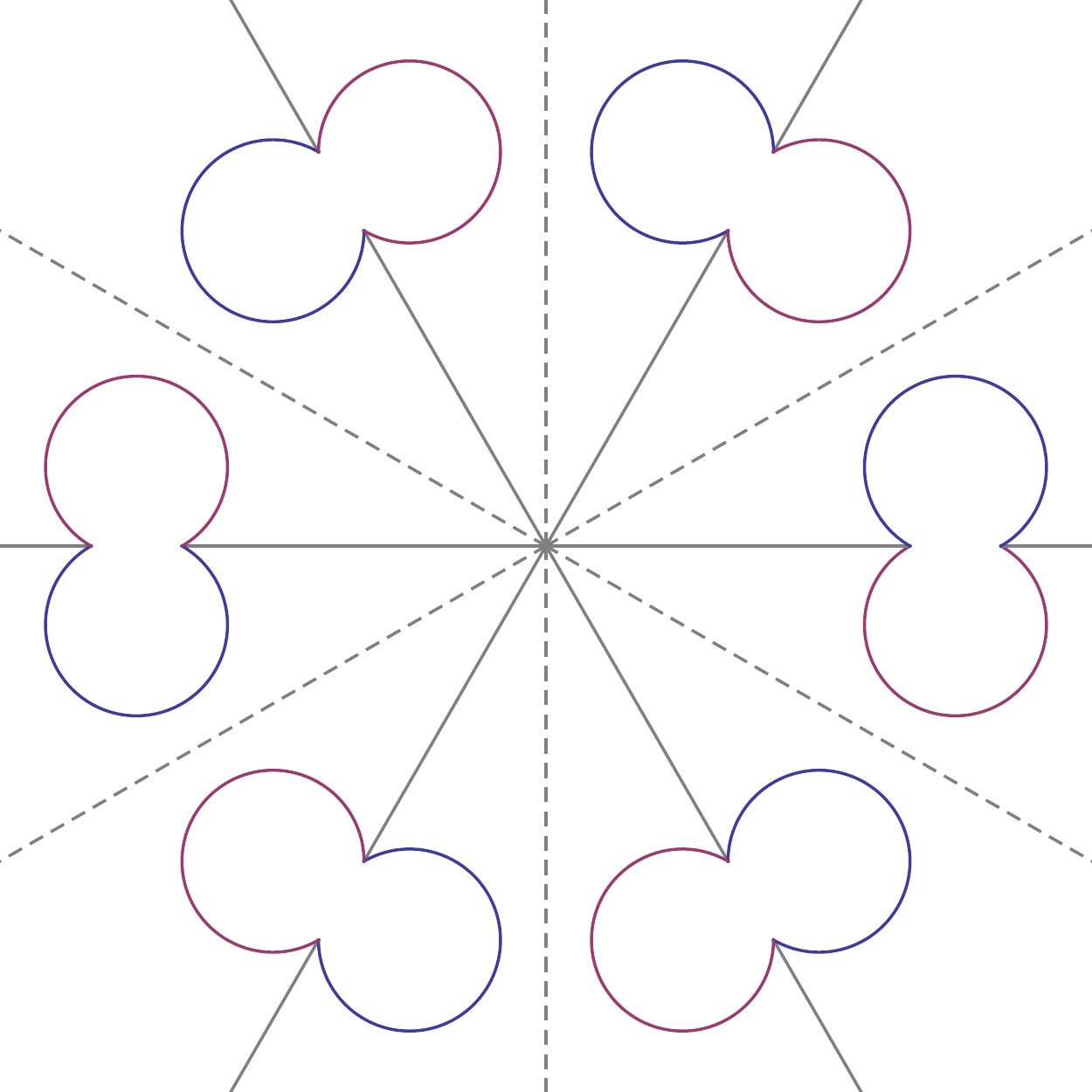}
\end{center}
\caption{ \label{eg6} The symmetric fundamental domain $\mathcal{D}_s$
for $n=6$.}
\end{figure}

\begin{figure}[h!]
\begin{center}
\includegraphics[scale=.5]{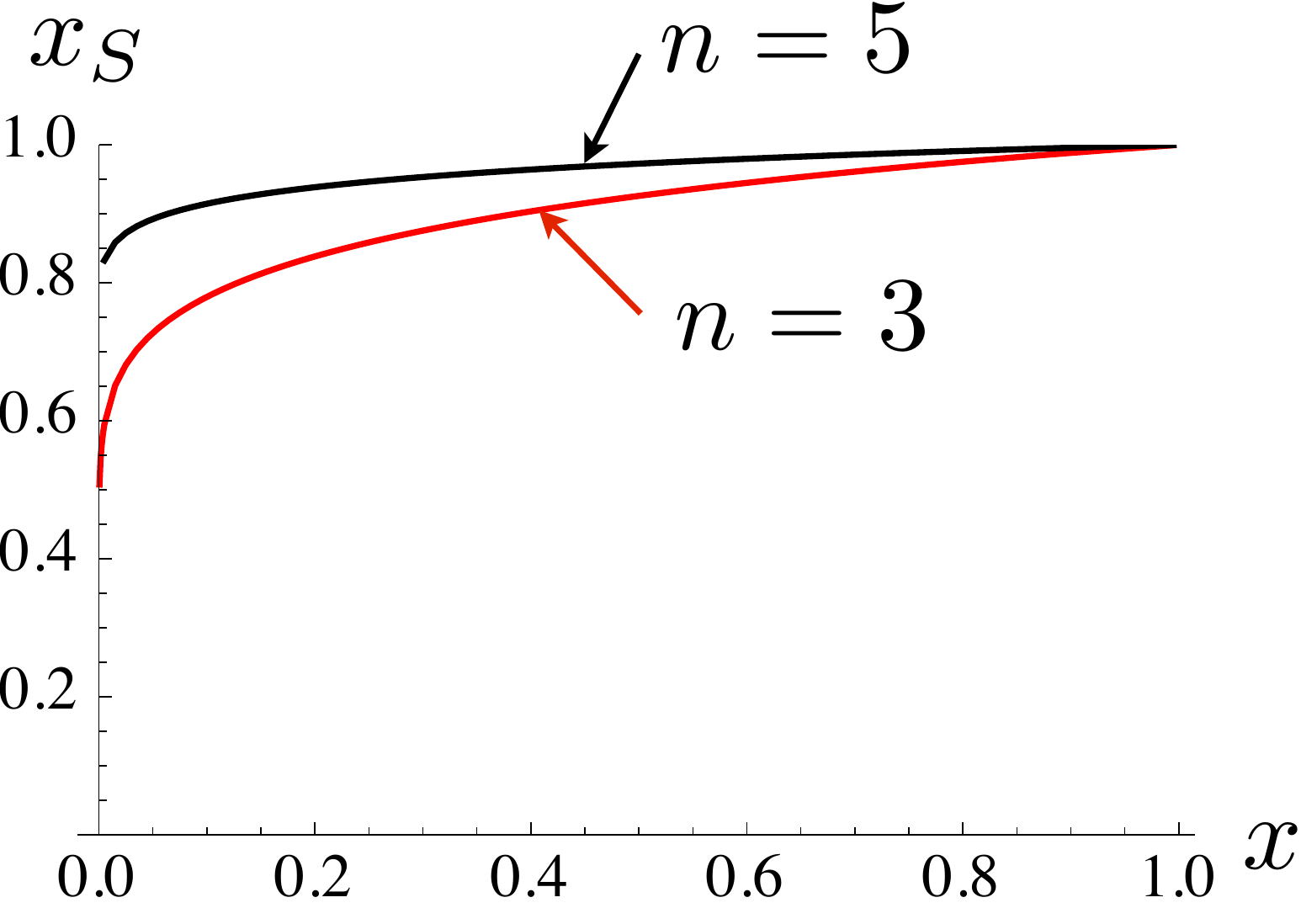}
\end{center}
\caption{ \label{xsx} The single parameter $x_S$ which goes into the Schottky
group for the surface $\mathcal{M}$ with monodromy conditions $\Gamma_\beta$.
For $x \rightarrow 0$ (where this saddle is subdominant to $\Gamma_\alpha$) the
limit is $x_S = cos^2(\pi/n)$ - although seeing this numerically requires high precision.}
\end{figure}

Note that the arcs start to cross unless we demand $\cos^2(\pi/n) < x_S$.
This actually corresponds to the boundary of moduli space $x=0$. See
Figire~\ref{xsx} for numerically calculated plots of $x_S(x)$.
The generators \eqref{gens} satisfy
\be
\Tr L_m = 2\frac{(x_S -  \cos(2 \pi/n)) }{1-x_S} 
\ee
and the condition that the arcs do not cross also corresponds to the requirement that the elements
are loxodromix  ($|\Tr L| > 2$.)  If we parameterize $L_m$ as:
\be
\frac{ L_m(w) -a_m}{L_m(w) - r_m} =q \left( \frac{ w - a_m}{w - r_m} \right)
\ee
then 
\be
q = \frac{ x_S - \cos(2\pi/n) - \sin(2 \pi/n) \sqrt{  (x_S/\cos^2(\pi/n)) -1} } 
{ x_S - \cos(2\pi/n) + \sin(2\pi/n) \sqrt{  (x_S/\cos^2(\pi/n))-1} } 
\ee
Note the parameter $q$ does not depend on $m$. The attractive
and repulsive fixed points $a_m,r_m$ are:
\be
|a_m| = |r_m|  = \sqrt{x_S} ,
\quad \arg(a_m,r_m) = \pi \frac{ ( 2 m -1)}{n} \pm \tan^{-1}\left(  \sqrt{  (x_S/\cos^2(\pi/n))-1} \right)
\ee
 The two fixed points come together at the boundary of moduli space
 when $\cos^2(\pi/n) = x_S$. 

The generators are not all independent: $L_{n} \ldots L_2 L_1 = (-1)^n$
where $-1$  acts trivially as a fractional linear transformation. 
This leaves $n-1$
generators for the group $\Sigma$. 
Which is the number  that we expected. 
The fundamental domain  is pictured in the left of Figure~\ref{fig:def} for $n=3$ and
Figure~\ref{eg6} for $n=6$ and uses the
real $z$-axis segments that we found above. This domain also does not conform to the standards of the usual fundamental domain of a Schottky group. This fact goes hand in hand with the over counting of generators. It is easy to see how to fix this. By deforming the circular arcs
in an appropriate way  the last generator $L_n$  becomes superfluous and one
is left with $2(n-1)$ identified closed circles rather than arcs. 
The argument is sketched in the right of Figure~\ref{fig:def}.
The existence of this deformed fundamental domain for the group generated by $L_1, \ldots L_{n-1}$ means that this group is by definition a classical Schottky group \cite{maskit}. 
In what follows we will go back and forth from considering these two 
different fundamental domains. We will refer to the first symmetric domain
as $\mathcal{D}_s$ and the later deformed domain as $\mathcal{D}_d$.

The boundary of the fundamental domain is defined as:
\be
\partial \mathcal{D}_s = \mathop{\bigcup}_{m=1}^n \left( U_m \cup \widetilde{U}_m \right)
\qquad \partial \mathcal{D}_d =  \mathop{\bigcup}_{m=1}^{n-1} \left( O_m \cup \widetilde{O}_m \right)
\ee
where $U_m, \widetilde{U}_m$ are the arcs given in \eqref{Um}, \eqref{Utm}
respectively. While the circles $O_m, \widetilde{O}_m$ are not uniquely defined; an
example is given in the right side of Figure~\ref{fig:def}.

The attractive and repulsive fixed points all lie along
the same circle $|w| = \sqrt{x_S}$ in the complex $w$ plane. This demonstrates a fact that was
speculated upon in \cite{Headrick:2012fk}. The authors showed that if the Schottky parameters $a_m,r_m, q_m$
can be chosen to be real then there is a so called real duality between the compact boson CFT
at the self dual radius and free fermions (with a fixed spin structure) and this
implied the EREs of these two theories were the same for two intervals.
They demonstrated the real duality using a different method and
speculated that this meant one could choose the Schottky parameters
to be real. Indeed we see here explicitly that this is the case - by making an $PSL(2,\mathbb{C})$
transformation to send the circle $|w| = \sqrt{x_S}$ to the real axis then
$a_m, r_m$ will all become real. $q$ remains fixed under this transformation and is real as above.
\footnote{ This also means that the Schottky group is actually a Fuchsian group
acting nicely on the disk $|w|<\sqrt{x_S}$. Interestingly this allows us to find a real-time three dimensional black hole
based upon $\mathcal{M}$, see \cite{Aminneborg:1997pz,Brill:1995jv} for details. These
are generalizations of the usual BTZ black holes \cite{Banados:1992wn}. It would be interesting to understand
what such a solution means for the EREs. }

Note that this reality argument applies to the Schottky parameters for our specific choice of A-cycles (cycles with trivial monodromy). It probably does not apply
for other replica symmetry breaking choice of A-cycles, although
for the arguments in \cite{Headrick:2012fk} one only needs the reality condition for
one such choice of cycles.

\begin{figure}[h!]
\label{bulkfund}
\begin{center}
\includegraphics[scale=.5]{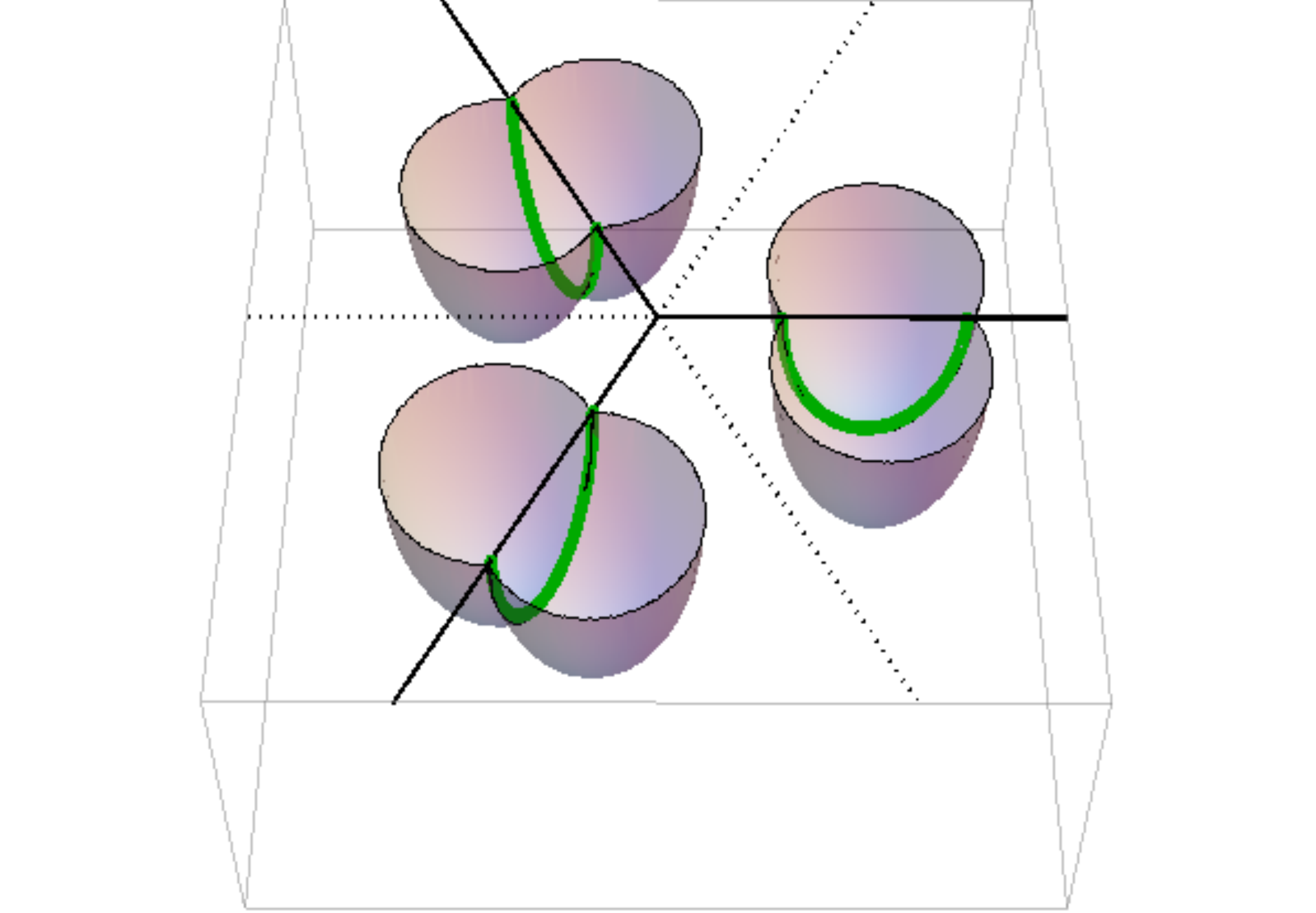}
\caption{ The gravity solution found by extending the arcs
$U_m, \widetilde{U}_m$ which live on the boundary of $AdS_3$ into hemispheres
inside the bulk and identifying these hemispheres. We show again the case $n=3$.
The picture also represents a (non standard) fundamental domain
for the quotient of $AdS_3$ by the action of $\Sigma$ which acts 
as \eqref{actionads3}. The hemispheres intersect over the green curves
which are bulk geodesics.
\label{fig:bulk}}
\end{center}
\end{figure}

Finally a picture of the bulk solution can be drawn by extending the
circles into hemispheres in $AdS_3$ \eqref{ads3}. We depict  in Figure~\ref{fig:bulk} 
the symmetric case where  the fundamental domain on the boundary is $\mathcal{D}_s$.
In this
picture two bulk hemispheres intersect over a geodesic.
We speculate that one can  identify this with a generalized version of geodesic
in the RT prescription. 

\section{Bulk action}

We will consider two ways to calculate the regularized on shell  Einstein action
for the bulk solution. Certain results in the literature will be used heavily.
To begin with we work on the deformed domain $\mathcal{D}_d$ defined in 
the previous subsection. It
was shown in \cite{Krasnov:2000zq} and further in \cite{tt} that  the on-shell 
action can 
be written in terms of a certain two dimensional Liouville action living on the 
boundary.
The action is defined on the domain
$\mathcal{D}_d$. It was first written down in \cite{zta} by Zograf and Takhtajan (ZT) and we will
refer to it as the ZT action. It was further shown in \cite{zta} that the variation of the ZT action
with respect to the moduli of $\mathcal{M}$ behaves nicely and we will use these
results to prove the assertion of the prescription that the variation of the EREs 
gives the accessory parameters \eqref{toint}.

After this we will go through a re-derivation
of the results in  \cite{Krasnov:2000zq} for the bulk action using the symmetric
fundamental domain $\mathcal{D}_s$ which is somewhat 
more convenient for our purposes. This will allow us to give an absolute
expression for the EREs not involving derivatives with respect to $z_i$. 

\subsection{Zograf-Takhtajan Action}

Firstly we  introduce the notion of Fuchsian uniformization.
We only need it as an intermediate step
so we will be brief.
This is another kind of uniformization compared
to the Schottky variety which aims to place a constant negative (for genus $(n-1) > 1$)
curvature metric on $\mathcal{M}$. The method is very similar to the Schottky
case. Consider the Poinc\'are disc $D$ with metric:
\be
\label{dshat}
d\hat{s}^2 = \frac{ dt d\bar{t} }{(1- |t|^2)^2}
\ee
where $|t| <1$. Fuchsian uniformization represents $\mathcal{M}$ as a quotient of $D$ by a discrete group which acts nicely on it. That is a discrete subgroup $\Sigma_F$ of $SL(2,\mathbb{C})$ which leaves the metric \eqref{dshat} invariant.  

Several results in the literature are available for computing the gravity partition function
when the metric is taken to be $d\hat{s}^2$. Because of the conformal anomaly the
result does depend on which metric we use within a fixed conformal class. 
We actually want the partition function on $d s^2$ given in \eqref{ds2}  and the
difference between these two is given by the Liouville action:
\be
\label{li}
Z_{\mathcal M}[ds^2 ] = e^{S_L} Z_{\mathcal M}[ d\hat{s}^2]  
\, , \qquad S_L = \frac{c}{96 \pi} \int_{\mathcal M} d^2 t \left( (\partial_t \phi_F)^2 - \frac{ 16  \phi_F}{(1 - |t|^2)^2}
\right)
\ee
where $\phi_F$ is the conformal factor which relates the two metrics:
\be
d\hat{s}^2 = e^{-\phi_F} ds^2  = e^{-\phi_F(z)} d z d \bar{z}
\ee
The metric $ds^2$ has conical singularities which means $\phi_F$ is singular at these points.
We deal with this by cutting out holes around these points which is a standard \cite{Lunin:2000yv}
procedure. Details are given in Appendix~\ref{sec:cut}.
To find $\phi_F$ we need to map the branched covering to this representation of $\mathcal{M}$.
Once again the ode \eqref{fuchs} allows us to construct the analytic
map and the quotient group $\Sigma_F$ as the monodromy group:
\be
\label{odefuchs}
t(z) = \frac{\psi_1(z)}{\psi_2(z)} \,, \qquad \psi'' + \frac{1}{2} T_{zz}^F  \psi = 0
\ee
where the stress tensor $T_{zz}^F$ has the same form as the Schottky case \eqref{fuchs} however now the accessory parameters $p_i \rightarrow p_i^F$ will be different. 
In order to fix the $p_i^F$ we must impose the condition that all monodromy elements generated by the fundamental group leave the metric \eqref{dshat} invariant. This is called
the Fuchsian monodromy condition. Note there is a unique condition here, we do not need to pick different ``A cycles'' and ``B cycles''. We emphasize that this is a different monodromy problem to the one given in  Section~\ref{sec:pre} and that we expect to find a different accessory parameter. 
The  field  $\phi_F$ is then:
\be
\phi_F = -  \ln | t'(z)|^2 + 2 \ln{(1 - |t(z)|^2)}
\ee
and we are now in a position to compute $S_L$. 

To calculate $Z_{\mathcal M}(d\hat s^2)$ we introduce the ZT action which
also happens to be a Liouville  type action however now living on the Schottky $w$ space.
Firstly one introduces a new Liouville field $\phi_S$ which is the conformal factor on the 
Schottky $w$-space which uniformizes that space, placing on it  a constant
negative curvature metric:
\begin{align}
\label{hypw}
 d\hat{s}^2 &\equiv e^{-\phi_S} d w d \bar w  \\
 \phi_S &=  -  \ln | t'(w)|^2 
+ 2 \ln{(1 - |t(w)|^2)}\end{align}
  
A major difference between the two Liouville fields is that $\phi_S$ is
not single valued on $\mathcal{M}$ where as $\phi_F$ is. 
As we move around on $\mathcal{M}$ the $w$ coordinate undergoes $PSL(2,\mathbb{C})$ transformations
along the B-cycles: $\widetilde{w} = L_m(w)$. In order that $\phi_S$ is consistent with these jumps in $w$ (the action of $\Sigma$) we must also require that $\phi_S$ jumps:
\be
\label{bc}
\phi_S(\widetilde{w}) = \phi_S(w) + \log | L_m'(w)|^2
\ee
under which the metric \eqref{hypw} is preserved:
\be
e^{-\phi_S(\widetilde{w})} d\widetilde{w} d \bar{\widetilde{w}} =  e^{-\phi_S(w)} d w d \bar w
\ee
The field $\phi_S$ is uniquely specified
by the identifications (or boundary conditions) and the requirement that it satisfies
the Liouville equation:
\be
\partial_{\bar{w}} \partial_w \phi_S = - 2  e^{-\phi_S} 
\ee
This equation follows from the requirement that \eqref{hypw} has constant negative curvature.
More succinctly $\phi_S$ is the solution of the equations of motion which follow from varying the following ZT action defined in \cite{zta}:
\begin{align}
\label{ls}
S_{ZT} &=  \frac{c}{96\pi} \int_{\mathcal{D}_d} d^2  w \left( \left( { \partial} \phi_S \right)^2 + 16  e^{- \phi_S} \right) + S_{ZT}^{bd} \\
\frac{96\pi}{c}  S^{bd}_{ZT}  
 & =     \sum_{m=1}^{n-1} \int_{O_m} ( 2 \phi_S  + \log |L_m'|^2+ 2  \log |c_m|^2 )\left( i d w \frac{L_m''}{L_m'} - i  d \bar{w} \frac{\bar{L}_m''}{\bar{L}_m'} 
\right) \\
&  \qquad  - 8 
\left( \lim_{R_w \rightarrow \infty} \int_{|w| = R_w} d\theta \phi_S    -  4 \pi \log R_w\right) \nonumber
\end{align}
where the boundary terms are  designed to impose \eqref{bc}.
Note that $c_m$ is the lower left component of the $L_m$ matrix defined
in \eqref{gens}.
Recall that $\partial \mathcal{D}_d = \cup_m (O_m \cup \widetilde{O}_m)$ where $\widetilde{O}_m = L_m(O_m)$.
Also note that the addition of the boundary term at $|w| = R_w \rightarrow \infty$ is to deal with an IR divergence due to the $w$ plane being infinite.  Typically one picks the generators $L_m$ using the freedom to conjugate by $PSL(2,\mathbb{C})$ such
that one of the fixed points is at $\infty$. Then this IR divergence is absent since  $w =\infty$
does not appear in the fundamental domain. This will not be convenient for us and we choose 
instead to directly deal with the IR divergence

In \cite{Krasnov:2000zq} it was shown  that the on shell value of the action \eqref{ls} for $\phi_S$
gives the regularized action of the bulk gravity solution (up to some minor additions, see 
\eqref{kras} below.)
The ZT action captures
the conformal anomaly and depends on the choice of metric 
in a fixed conformal class. 
The appropriate
metric here is \eqref{hypw} and by the
usual dictionary of AdS/CFT the action $S_{ZT}$ gives a contribution to the partition function 
$Z_{\mathcal M}(d \hat{s}^2)$ of the CFT defined on this metric:
\be
\label{kras}
\hat{S}_{\rm gr}^\gamma   = - S_{ZT} + \frac{c}{2} (n-2)   -  \frac{c}{3} (n-2) \log \Lambda
\ee
where $\gamma$ labels the particular gravitational saddle. We have
also included a UV cutoff factor $\propto \log \Lambda$  which cannot
be removed in the limit $\Lambda \rightarrow 0$ due
to the conformal anomaly. It is proportional to the Euler character of $\mathcal{M}$
which is $-2(n-2)$. 

Putting everything together the partition function we seek  \eqref{sadsum} is the sum
over the different saddles $\gamma$ (including the ones we do not construct):
\be
\label{newsadsum}
Z_{\mathcal M} (d s^2) = \sum_{\gamma} \exp( - \hat{S}_{\rm gr}^\gamma + S_L + \mathcal{O}(c^0) )
\ee

Rather than work directly with $S_{ZT}$ and $S_L$ we would like to compute their on-shell variation with
respect to the $z_i$. We can use several results in the literature. These results
can be understood as essentially arising from conformal ward identities.

For $S_L$ the variation was given originally by Polyakov using
the Liouville theory path integral. A proof using just the classical Liouville action
was given in \cite{ztb,ztc,Cantini:2001wr,Hadasz:2003kp}. The variation gives the
\emph{Fuchsian} accessory parameters defined in terms of the Fuchsian
monodromy problem: \footnote{In the literature on Liouville theory one
considers a slightly different form of the Liouville action $S_L$ from \eqref{li}. 
Firstly the action is defined on the $z$ plane with reference metric $ds^2$. We can get to this
form by integrating by parts on \eqref{li} and we go through this in Appendix~\ref{sec:cut}.
Secondly the $z$-plane is not multi sheeted, rather the points $z_i$
are conical deficit singularities for the uniform metric $d\hat{s}^2$.
One can think of these as arising from a quotient
of our surface $\mathcal{M}$ by $\mathbb{Z}_n$. This explains the extra factor of $n$
in \eqref{Lvar} compared to for example \cite{Hadasz:2003kp}. }
\be
\label{Lvar}
\frac{\partial S_L}{\partial  z_i} = \frac{c n }{6} p_i^F
\ee

For $S_{ZT}$ the results in \cite{zta} and \cite{Hadasz:2003kp} can be applied. 
Here a mathematically rigorous procedure for varying the
action with respect to the moduli was used and goes under the name of quasiconformal transformations.
From the results in these papers we can derive:
\be
\label{ZTvar}
\frac{\partial S_{ZT}}{\partial z_i} = \frac{c n}{6}  (p_i^\gamma  - p_i^F) 
\ee
the difference in Fuchsian and Schottky accessory parameters. 
Details are given in Appendix~\ref{sec:quasi}. The reason
the difference in accessory parameters appears is that 
the stress tensor of $S_{ZT}$ which appear when varying the action takes the form:
\be
\widehat{T}_{ww} =  \partial_w^2 \phi_S + \frac{1}{2} (\partial_ w \phi_S)^2
 = w'(z)^{-2} \left( T_{zz} - T^F_{zz} \right)
  = w'(z)^{-2} \left( \sum_i \frac{(p_i^\gamma - p_i^F)}{z-z_i} \right)
\ee
This stress tensor is the one associated to the ground state
of the CFT living on $d\hat{s}^2$.

Adding \eqref{Lvar} to \eqref{ZTvar}  explains equation \eqref{toint} given in the prescription of Section~\ref{sec:pre} which was the main goal of this current subsection.
Note that $p_i^F$ cancels  between  \eqref{Lvar} and \eqref{ZTvar} so the final result does not depend on the Fuchsian uniformization. This is expected since we used
the uniform metric $d \hat{s}^2$ only as an intermediary.

\subsection{Regularization Surface}

The goal of the next two subsections is to find an expression for $S_{gr}$ without
resorting to taking derivatives thereof. To do this we go through
the derivation in \cite{Krasnov:2000zq} using a slightly different regularization
procedure.

Following  \cite{Krasnov:2000zq} we need to pick a regularization surface in order to define the bulk action.
This surface should be consistent with the $L_m$ identifications.
Also the desired boundary metric  should
be induced on this cutoff surface . We use the
field $\phi_S$ to define our cutoff surface:
\be
\xi \approx \Lambda e^{\phi_S/2}  \qquad ds^2|_{\Lambda}  \approx \Lambda^{-2} e^{- \phi_S} 
d w d \bar{w} = \Lambda^{-2} d \hat{s}^2
\ee
with UV cutoff $\Lambda \rightarrow 0$. Note we could not simply cutoff at fixed $\xi$ since
under the $PSL(2,\mathbb{C})$ isometries \eqref{actionads3} the coordinate $\xi$ changes.
For small $\xi$ this transformation is consistent with the transformation
of $e^{\phi_S/2}$. Hence this choice.  Using this surface  which we call $\mathcal{D}_\Lambda$ we define the regularized action in the usual way:
\be
\label{sgr}
16 \pi G_N \hat{S}_{gr} = \int_{Q_\Lambda}d^3 x \sqrt{g} ( R  - 2 ) + 2 \int_{\mathcal{D}_\Lambda} d^2 x \sqrt{h}
(1- K )
\ee
where $Q_{\Lambda}$ is the regularized portion of the fundamental domain for the quotient $AdS_3'/\Sigma$  with boundary $\mathcal{D}_\Lambda$. $K$ is the trace of the extrinsic curvature
and $h$ is the induced metric. Note that depending on the choice of
fundamental domain for the quotient $Q_{\Lambda}$ the boundary 
will be conformally 
equivalent to either $\mathcal{D}_s$ or $\mathcal{D}_d$ as $\Lambda \rightarrow 0$.
The choice of $\mathcal{D}_d$ was worked out in \cite{Krasnov:2000zq} so here
we pick $\mathcal{D}_s$. 

We can be a little more precise and
pick coordinates $\{ \Lambda, u, \bar{u} \}$ in order to write the original
$AdS_3$ metric in the Fefferman-Graham expansion with induced metric $d \hat{s}^2$.
Following \cite{Krasnov:2001cu} write
\be
\label{fg}
\xi = \frac{ 4 \Lambda e^{\phi_S/2}}{ 4 + \Lambda^2 e^{\phi_S} | \partial_u \phi_S |^2 }
\qquad w  = u - \partial_{\bar u} \phi_S  \frac{ 2 \Lambda^2 e^{\phi_S}}{ 4 + \Lambda^2 e^{\phi_S} | \partial_u \phi_S |^2 }
\ee
where now $\phi_S$ is considered a function of $\phi_S (u,\bar{u})$ which anyway
approaches $(w,\bar{w})$ at the boundary. The bulk metric is:
\be
\label{fgmet}
ds^2 = \frac{ d\Lambda^2}{\Lambda^2} + \frac{1}{\Lambda^2} \left| \left( \frac{ 1 + \Lambda^2 }{ 1 - |t|^2} \right) d\bar{t} 
+ \frac{1}{2} \Lambda^2 (1 - |t|^2)  \hat{T}_{tt} d t \, \right|^2
\ee
where we have written the answer in terms of the Poinc\'are disk coordinate $t$.

The answer has a particularly simple form since the FG expansion terminates
in three bulk dimensions \cite{Skenderis:1999nb,Banados:1998gg}.
The stress tensor $\hat{T}_{tt}$ of the field theory living on $d\hat{s}^2$ 
and in the state defined by the saddle at hand appears
as a sub leading term in the FG expansion \cite{Balasubramanian:1999re}.
In these coordinates we can write the stress tensor as\footnote{
The normalized stress tensor of the state is related to this by a factor
$\hat{T}_{CFT} =  \frac{c}{24 \pi} \hat{T}$. 
}:
\be
\hat{T}_{tt} =( t'(z) )^{-2} \left( \left\{ w,z \right\} -  \left\{ t,z \right\} \right)  = 
( t'(z) )^{-2} \left(T_{zz} - T_{zz}^F \right)
\ee
where $t,w$ were given in terms of solutions to the appropriate Fuchsian \eqref{odefuchs}
or Schottky \eqref{ratsoln} odes . 
We use the notation
$\hat{T}$ to denote the stress tensor for the vacuum state of the theory living
on the uniform metric $d \hat{s}^2$. While $T$ is reserved for the stress tensor of the theory
defined on the singular metric $d s^2$.

Fefferman-Graham coordinates typically develop a coordinate singularity away
from the boundary. This is indeed the case, the metric becomes degenerate
when:
\be
\Lambda^2_c 
 = \frac{4}{ (r_n^2 - |t|^2) | \hat{T}_{tt}| - 4 r_n^2 } 
\ee

Actually we could attempt to bypass altogether the Fuchsian uniformization - and never
even mention $d\hat{s}^2$ or the Poinc\'are disk coordinates $t$. 
In this case we should pick our regularization
surface such that  the induced metric is directly $d s^2$ which is anyhow the desired
metric. The fact that the partition function depends on this regularization
surface in the limit $\Lambda \rightarrow 0$ is a manifestation of the Weyl
anomaly in holography \cite{Henningson:1998gx}.
 This would leave the introduction of the Liouville action $S_L$ unnecessary. 
We can do this by setting 
\be
\label{phi}
\phi \equiv \phi_S - \phi_F = \ln w'(z) + \ln \bar{w}'(\bar{z})
\ee
and using $\phi$ to define a new regularization surface. 
Note that $\phi$ is a locally harmonic function on the $z$ plane.
Since $\phi_F$ is single valued on $\mathcal{M}$ the new field transforms in the same way  as $\phi_S$ around the non-trivial B-cycles \eqref{bc}.  

We can pick Fefferman-Graham coordinates with respect to the Liouville
field $\phi$ by replacing $\phi_S \rightarrow \phi$ in \eqref{fg}. The bulk metric is then:
\be
ds^2 = \frac{ d\Lambda^2}{\Lambda^2} + \frac{1}{\Lambda^2} \left| d \bar{z} + \frac{\Lambda^2}{2} T_{z z} dz
\right|^2
\ee
From
this we see that $T_{zz}$ is the stress tensor of the theory living on $ds^2$. 
This is stress tensor that appears in the original ode.

Compared to the Fefferman-Graham coordinates
for the metric $d\hat{s}^2$ those for $d\hat{s}^2$ are rather singular.
Here we find a breakdown of the FG coordinates arbitrarily close to the points on the boundary $(z \rightarrow z_i, \Lambda \rightarrow 0)$. This breakdown was discussed
and confronted in \cite{Hung:2011nu} in a similar computation of EREs. 
They break down at $\Lambda_c^2 = 2/|T_{zz}| \sim |z-z_i|^2$.
This is because in addition to UV regulating the theory using the cutoff
surface $\Lambda=$ constant we  need to regulate the divergences
associated with the conical singularities in $ds^2$. We achieved this previously 
by using the singular Liouville field $\phi_F$ to transform to the non-singular
metric $d\hat{s}^2$. Then the Liouville action for $\phi_F$ contained the divergences
associated to these conical singularities.  So the field $\phi$ has to take
into account both the divergences associated to the conical singularities as well as the $L_m$ identifications. 
We found it convenient to deal with these issues separately by splitting this
into two steps.

\subsection{Action from the Symmetric Domain}

We are now ready to calculate \eqref{sgr}. 
Firstly let us compute the bulk integral  using the $(\xi,w,\bar{w})$ coordinates.
We use the FG coordinate \eqref{fg} to define the regulating surface at $\Lambda = $ const. 
The fundamental domain was depicted in Figure~\ref{fig:bulk} consisting of removing
hemispheres from $AdS_3$. Define $V$ the volume of this domain:
\be
\label{eh}
  \int d^3 x \sqrt{g} (R-2) = 4V =  4 \int d^2 w \int_{\xi_{\min}}^{\infty} \frac{d \xi}{\xi^3} = 2 \int d^2 w \frac{1}{\xi_{\min}^2(w,\bar{w})}
\ee
Away from the hemispheres the radial integral is cutoff at
$\xi_{\min}^{-2} = \Lambda^{-2} e^{-\phi_S(w,\bar{w})} -  | \partial_w \phi_S|^2/2
+ \ldots$ where we should emphasize that we are working with $w,\bar{w}$ coordinates
at the boundary and not $u,\bar{u}$. We define $V_m,\widetilde{V}_m$ the volumes of the chunks of $AdS_3$ below the hemispheres segments  (below in the sense of Figure~\ref{fig:bulk}).
There are $2n$ of these but by the replica symmetry they are all the same:
\be
 V =  \frac{1}{4} \int_{\mathcal{D}_s} d^2 w \left( 2 \Lambda^{-2} e^{-\phi_S} -  | \partial_w \phi_S |^2
\right) + 2 n V_1
\ee
Where $V_1$ corresponds to the ``first'' hemisphere segment - on the
boundary it becomes the $m=1$ segment of \eqref{Um}.
The volume is:
\be
V_1 = \frac{1}{2} \int_0^{2\pi} d \theta \int^{r_m(\theta)}_0 r dr \frac{1}{\rho^2 - r^2}
\,,\qquad \rho = \frac{(1-x_S)}{2 \sin(\pi/n)}
\ee
where we do the integral using cylindrical coordinates about
the center of the hemisphere. We have given the radius of the hemispheres  $\rho$
in terms of the Schottky parameter $x_S$ defined in Section~\ref{sec:bs}. 

Examining the geometry of the hemispheres shown in Figure~\ref{fig:bulk} we see
that we get two terms, one from where the $r$ integral is cutoff by the intersection
of the hemisphere with the regularization surface and the other from the
remaining triangular shaped region. That is where the radial
integral is cutoff by the intersection with another hemisphere. These two terms are:
\footnote{The expression for $N_1$ is only a function of $n$ and does not analytical continue well to $ n <2$. We guess an expression that has a better continuation in $n$
\be
N_1 = - \frac{1}{2} {\rm sign} (n-2) \int_0^{ \min(\pi/n, \pi - \pi/n)}  d \theta \ln \left( 1 - \frac{ \cos^2(\pi/n)}{\cos^2(\theta)} \right) 
\ee
For $n<2$ this expression \emph{subtracts} the volume of a triangular shaped region,
since now the other term in $V_1$ over count the volumes of the hemisphere segments. This is  a guess since the bulk solution does not make any sense for $n<2$. This guess seems
to yield the correct answer. 
}
\be
\label{volume1}
V_1 =  - \frac{1}{4} \int_{U_1} d\theta \left( \phi_S + 2 \ln(\Lambda/\rho) \right)
+ N_1 \,, \quad N_1  = - \frac{1}{2}  \int_0^{\pi/n} d\theta \ln \left( 1 - \frac{ \cos^2(\pi/n)}{\cos^2(\theta)} \right) 
\ee
The first term is an integral on the $AdS_3$ boundary along the segment
in the $w$-plane which can be described as (see \eqref{Um}):
\be
U_1 = \left\{ w = w_1+ \rho e^{i \theta} 
\,\,; \, - \frac{\pi}{2} +  \frac{ \pi}{n} < \theta < \frac{3\pi}{2}-   \frac{ \pi}{n}  \right\}\,, \qquad
w_1 = \left( \frac{ x_S - e^{- i 2\pi/n}}{ 1 - e^{- i 2\pi/n}}\right)
\ee
where $w_1$ is the center of the circular arc in the complex plane.
Note that along this arc $\phi_S$ is identified under $L_1$ with the $\phi_S$ at the next
arc moving in an anti-clockwise direction on Figure~\ref{fig:bulk}. 
We can write this identification \eqref{bc} simply as:
\be
\phi_S\left(\ \widetilde{w}_1 + r^{-1} \exp(- i \theta + i2 \pi /n ) \right)
 = \phi_S\left(w_1 + r  \exp( i \theta) \right) \, -  4 \ln(r/\rho)
\ee
where $\widetilde{w}_1$ is the center of this adjacent arc ($ \widetilde{w}_1 = e^{i 2\pi/n} \bar{w}_1$).

An issue we have ignored so far is related to the IR divergence associated
with working in the Poincare patch.  To fix this we momentarily move to 
global coordinates where the metric is:
\be
ds^2 = \frac{ d\xi^2}{\xi^2} + \left( \frac{R_w}{\xi} - \frac{\xi}{R_w} \right)^2 R_w^2
\frac{ dw d\bar{w} }{ (R_w^2 + |w|^2)^2} 
\ee
and where the radial coordinate ranges over $0 < \xi < R_w$.
The limit $R_w \rightarrow \infty$ returns us to Poincare coordinates.
However before we take this limit we get an extra log contribution to 
to the bulk Einstein action:
\be
 V  
 =  \int d^2w  \frac{R_w^2}{(R_w^2 + |w|^2)^2} \left( \frac{R_w^2}{2 \xi_{\min}^2}
 + 2 \log( \xi_{\min}/R_w) + \ldots \right) 
\ee
The log term encodes the coupling of $\phi_S$ to the curvature
of the $w$ sphere, which we have hidden at $|w| \rightarrow \infty$ by
working on the plane. We must keep this term which 
in the limit $R_w \rightarrow \infty$ gives us the addition:
\be
 V \rightarrow  V + \frac{1}{2} \int_{|w|=R_w} d \theta \phi_S   -  2 \pi \ln(R_w/\Lambda)
\ee

The extrinsic curvature part of the gravitational action \eqref{sgr} is most conveniently evaluated
in FG coordinates $(\Lambda,u,\bar{u})$ - which can be related to $(\xi,w,\bar{w})$ coordinates
close to the boundary (note if we were using $\phi$ and not $\phi_S$ as our Liouville field there would be some extra complications to deal with here.) That is:
\be
2 \int d^2 x \sqrt{h} ( 1 - K)
 \approx 
 - \int d^2 w \left(  2 e^{-\phi_S(w,\bar{w})} \Lambda^{-2}
 + 4   \partial_w \partial_{\bar w} \phi_S  \right)
\ee
Adding everything together the quadratic UV divergence associated to $\Lambda$ vanishes
leaving:
\be
\frac{96 \pi}{c}  \hat{S}_{gr}  = - \int_{\mathcal{D}_s} d^2 w \left( (\partial \phi_S)^2 + 4 \partial^2 \phi_S 
\right)   +  32 n V_1   + 8 \int_{|w| = R_w} d \theta \phi_S - 32 \pi \ln( R_w/\Lambda) 
\ee
where one should remove $|w| > R_w$ to define the symmetric domain $\mathcal{D}_s$.
The $\partial^2 \phi_S$ term evaluates to something proportional
to the Euler character of $\mathcal{M}$:
\be
 \int d^2 w  \partial^2 \phi_S  = - 8 \pi(n-2)
\ee

Combining the gravitational action with the Liouville
action \eqref{li} as in \eqref{newsadsum} we find
after integrating by parts:
\bea
\nonumber
\frac{96 \pi}{c} S_{gr} 
& &= - \int_{\widehat{\mathcal{D}}_s }  d^2 w ( \partial \phi )^2
  - 8 n \int_{U_1} d \theta( \phi - 2 \ln \rho )
 + 8 \int_{|w| = R_w} d \theta \phi - 32 \pi \ln R_w    \\
 && \hspace{1cm}  + 32 \pi (n-2)  \ln \Lambda + 32 \pi (n-2)  + 32 n N_1 
 \label{inter}
\eea
where everything is written in terms of $\phi \equiv \phi_S - \phi_F$.
Recall that this is a harmonic field that can be defined solely in terms
of the Schottky uniformization coordinates $w(z)$ see \eqref{phi}. 
The new domain $\widehat{\mathcal{D}}_s$ is a  regularized 
version of $\mathcal{D}_s$ defined by
cutting out various holes where $\phi$ diverges.  
The justification
for this cutting procedure is the same as for the Liouville
action that we went through in Appendix~\ref{sec:cut}.
Some details have been swept under the rug in arriving at \eqref{inter}.
For example we need to disentangle the conical singularity at $z_4$ from
the curvature singularity of the infinite $w$ plane at $w = \infty$. Recall
that for our choice \eqref{sch} these points were the same $w(z_4) = \infty$.  
The quickest
way to deal with this is to deform
the point $w(z_4) \rightarrow w_4 \gg 1$ such that $w(z_\infty) = \infty$
for $z_\infty \approx z_4$ on only one of the replicas. Taking $w_4 \rightarrow \infty$
of this procedure defines the action \eqref{inter}. 

We can now evaluate \eqref{inter} since $\phi$ is harmonic. 
After some work (the details of which are given in Appendix~\ref{sec:cut2}) one finds an answer which can be
succinctly written in terms  of $\psi_-$ the particular solution to the ode 
appearing in the denominator of the Schottky coordinate $w = \lambda \psi_+ /\psi_-$ 
(see the discussion around \eqref{sch}.) We send $z_1 = 0, z_2 = x, z_3 =1$
and $z_4 \rightarrow \infty$. By defining the Mutual Information (for this
particular saddle $\gamma = \beta$) as in \eqref{midef} we can take the limit $z_4 \rightarrow \infty$ without the associated IR divergence.  We find:
\bea
\nonumber
\label{master}
c^{-1} I_n^\beta  &=& \frac{n}{12 \pi (n-1)} \int_{x}^{1} dz \Im \left( 
 \frac{\psi_-'}{\psi_-} - \frac{\widetilde{\psi}_-'}{\tilde{\psi}_-} \right) \ln \left| \frac{\psi_-'}{\psi_-}
-  \frac{\widetilde{\psi}_-'}{\tilde{\psi}_-}  \right|  - \frac{n N_1}{3\pi(n-1)} 
 \\
&&  \hspace{1cm} + \frac{1}{12} \ln \left| \frac{\mu_1 \mu_2 \mu_3}{\widehat{\mu}_4 \rho^2 } \right| + \frac{1}{6} \left( \frac{1}{n} + 1  \right)
\ln(x)   
- \frac{ (n+1)\ln(n)}{6 (n-1)}
\eea
where $\psi_- \approx z^{1/2-1/(2n)}$ at the origin of the $z$ plane. 
The integral is along the real axis  with $\psi_- = \psi_-( z + i\eta)$ evaluated just
above the real axis and  we have defined  $\psi_-$ evaluated just below the real axis
as $\widetilde{\psi}_- =\psi_-(z - i \eta)$.
Note that $\widetilde{\psi}_-$
is related to $\psi_-$ by a monodromy around the loop
on $\mathcal{M}$ which connects the top of the real line segment
$z \in \left[x,1 \right] + i \eta$ with the bottom $z \in \left[x,1\right] - i \eta$.
See the left panel in Figure~\ref{fig:zw}. 
This is the monodromy loop that defined the matrix $L_1$
and we can write  $\widetilde{\psi}_- = c_1 \psi_+ + d_1 \psi_-$ .
Recall that we are studying the saddle associated to the
monodromy conditions in  $\Gamma_\beta$. A similar
expression to \eqref{master} exists for $\Gamma_\alpha$.
Finally $\rho$ and $\mu_i$ are also extracted from $\psi_-$ simply as:
\be
\rho = \left| \frac{\lambda n^{-1}}{W[\psi_-,\widetilde{\psi}_-]}  \right|
\qquad \mu_i =  \lim_{ z \rightarrow z_i}  \frac{\lambda }{ \psi_-^2}
 (z-z_i)^{1-1/n} \quad \hat{\mu}_4 = - \lim_{z \rightarrow \infty}  \frac{\lambda}{
 \psi_-^2} z^{1-1/n}
\ee
where $W$ is the Wronskian of two solutions to the ode.
Note that when we plug these constants into 
\eqref{master} the factor $\lambda$ drops out so we can
effectively ignore it.

We have confirmed numerically that the expression \eqref{master} gives
the same answer  as the one obtained by integrating
the accessory parameter. 
We think formula like \eqref{master} should generalize for other
non-replica symmetric saddles which would be useful in checking the
assumption that replica symmetry remains unbroken. The details of course
will be slightly different and we leave this to future work.

\section{Discussion}

To summarize we have calculated some contributions to  EREs in holographic
CFTs by constructing higher genus gravitational handlebody solutions.
We only found a subset of all possible classical solutions.
These are the solutions which were highly symmetric  -
respecting the symmetries of the boundary surface.
We found that the bulk actions of this set of solutions  continued nicely 
under $n \rightarrow 1$ to the Ryu-Takayanagi formula involving lengths of
bulk geodesics. We did this for arbitrary numbers of intervals,
however  the bulk solutions were only described in detail for $N=2$.

If we could show that the symmetric solutions dominate in the sum over saddles
 at large central charge  for all $0<x<1$ then we would have found
the Mutual Renyi Information for all $1+1$ CFTs with an Einstein gravity dual description and
we would have proven the RT formula in this case. We have not managed to
come up with a proof necessary for this purpose and in fact after some thought
we are not sure it is true. 

More conservatively for two intervals one should be able to show that our
prescription for computing the EREs is correct  for any $n$ in a perturbative expansion about
$x=0$.  
We can argue pictorially that the saddle which we called $\Gamma_\alpha$ 
dominates over all other gravitational saddles in the limit $x\rightarrow 0$. 
In particular  the minimal length
of a curve living on $\mathcal{M}$ (in terms of either metric $ds^2$ or $d\hat{s}^2$)
which is homologous to the cycle $C_\alpha$ (see Figure~\ref{fig2cyc}) becomes parametrically small compared to the minimal length curve  homologous to $C_\beta$. This is true on all replicas.
The bulk solution which has the least volume and hence
least action will be the
one where all the short cycles $C_\alpha$ are contractable.
If any of the other longer $C_\beta$ cycles were contractible then we will clearly
get a larger volume subdominant solution.  

Once we have shown $\Gamma_\alpha$ is the dominant saddle
as $x \rightarrow 0$ then at infinite central charge the other saddles cannot
be seen to all orders in an expansion about $x=0$. This leaves
open the possibility that another non-symmetric solution becomes dominant
at some \emph{finite} value of $0<x<1/2$. The danger
region is $x \approx 1/2$ since then the minimal length curves of $C_\alpha$
and $C_\beta$  approach the same length.
Similar arguments can be made about the point $x=1$ as well as for
more than two intervals.

The paper \cite{hartman} found the same monodromy prescription that
we gave in a completely different manner.  They studied semi-classical (in the
sense of large central charge) Virasoro conformal blocks for low dimension operators which 
can be computed in terms of this monodromy condition \cite{recursion}. The ERE thought
of as a 4-point function of twist operators was then argued to receive its dominant contribution from the conformal block for exchange of the unit operator (including the stress tensor and its descendants.)  These conformal blocks can be identified with the bulk solutions
 we constructed and since they contain the unit operator in the s-channel (t-channel) exchange 
 as $x \rightarrow 0$ ($x\rightarrow 1$) 
 must give the dominant contribution to the 4-point function about $x=0$ ($x=1$). 
 The question of intervening saddles in the danger region
 was also unresolved in \cite{hartman} and would involve the 
 conformal block of some other heavy operator  (with conformal dimension $\mathcal{O}(c)$) potentially becoming dominant. 

We now review some material that may help for trying to construct these missing saddles
in this danger region.

\subsection{The missing saddles}

Summing over saddles in $AdS_3$ holography is a well studied
subject at genus one (see for example \cite{Maldacena:1998bw,Maloney:2007ud}) but is less explored at higher genus (however see \cite{Yin:2007gv,Witten:2007kt}.) The basic idea can be understood in terms of the moduli space of Riemann surfaces $M_n$. At genus $(n-1)$ this
is a $3(n-2)$ complex dimensional space. The covering space  of $M_n$ is called
Teichm\"uller space $T_n$ and this space distinguishes Riemann surfaces
related by large coordinate transformations - sometimes referred to as
Modular transformations or  elements in the Mapping Class Group (MCG). $M_n$
is then just the quotient of $T_n$ by the MCG. If one studies a modular invariant
CFT on a Riemann surface then the partition function is 
a function on $T_n$ which is consistent with the action
of the MCG - thus it is also well defined on $M_n$. 

The handlebody solutions to $AdS_3$ gravity 
however are not invariant under the action of the MCG. This is because we had to choose a
set of $(n-1)$ A-cycles which were contractable in the bulk
and these cycles change under the MCG action. This means
that the gravitational action thought of as a function
on $T_n$ becomes multivalued on the quotient $M_n$.  
To find a modular 
invariant partition function we need to sum over all images of the MCG
\footnote{not all!, a subgroup leaves the bulk solution untouched, but
we do not dwell here on such details.}
Schematically,
\be
Z_{\mathcal M} = \sum_{g \in MCG} \exp \left( - S_{gr}^{ g(\alpha)} + \mathcal{O}(c^0) \right) 
\ee
where we continue to use $\alpha$ to denote the replica symmetric solution corresponding
to the monodromy conditions specified by $\Gamma_\alpha$. The action $g$ on
this solution which we  denote $g(\alpha)$ then
scans through  all relevant handlebody solutions.
After the sum $Z_\mathcal{M}$ is modular invariant and well defined
on $M_n$.

For example there is one element
in the MCG which sends $\Gamma_\alpha \rightarrow \Gamma_\beta$
. The phase transition between these two saddles at $x=1/2$ is  a fixed point of this
action. The other solutions $g(\alpha)$ have not yet been constructed.
At finite central charge $c$ we would need all of them to find a consistent partition
function, however at large $c$ we can ignore all but the dominant ones.
One then just needs to find which is the dominant  as a function of $x$.

We have found $S_{gr}$ along a one dimensional slice in $T_n$ 
which is the special slice distinguished by the replica symmetry. 
We just need to construct the other one dimensional slices of $T_n$ related
by elements of the MCG to the replica symmetric slice. Since the replica
symmetry acts non-trivially on these solutions there will be more than
one slice of $T_n$ space (related by the replica symmetry) with the same action.
If one of these is dominant then it will clearly give rise to the phenomena that
might be called replica symmetry breaking. 
Actually we require infinite central charge otherwise the usual arguments about
the lack of symmetry breaking in a finite system apply.
It would be interesting to understand
this phenomenon (if it were to occur) in more detail. 
Some questions that come to mind: What is the order parameter?
What does this imply about the spectrum of the reduced density matrix?
What are the implications for the product orbifold theory and EREs written as twist
operator correlation functions?\footnote{We give an answer to the first question in this footnote:
the non-symmetric accessory parameters $p^s$ (for $s\neq 1$) are good order parameters,
and can be extracted from integrals of stress tensor on $\mathcal{M}$, see \eqref{stress}. Formally one would need to slightly break the replica symmetry in order to get a non-zero
expectation value for this order parameter. Then at large central charge one can
remove the symmetry breaking term and still potentially arrive at
a non-zero answer for $\<p^s\>$, indicating symmetry breaking has occurred.  }

Unfortunately finding the other solutions
is easier said than done. We sketch how one might construct these
missing saddles using the methods of this paper. Pick
a non-symmetric set of $(n-1)$ A-cycles and demand that they are contractible in the bulk.
The ode \eqref{fuchs} can be used to construct the bulk solution - however
as we alluded to throughout the paper  we need to turn on the other accessory parameter
 \eqref{stress}. Which ones we have to turn on depends on the
symmetry breaking pattern.
Already we see the problem becomes intractable for large $n$
- we have to search in the $3(n-2)$ dimensional
space of accessory parameters in order to satisfy the monodromy conditions.
However it is feasible to attempt this for small values of $n$. One then
needs to compute $S_{gr}^{g(\alpha)}$. For this we can no longer resort
to integrating the accessory parameters and instead have to work with
an absolute expression  for $S_{gr}$. This is where a suitably
generalized version of \eqref{master} will  come in handy.

To further complicate things there are nonhandlebody solutions
when $n>2$. See \cite{Yin:2007at} for a discussion in the context
of partition functions of $AdS_3$ gravity. See also \cite{tt}
where these solutions are constructed
using Quasi-Fuchsian Kleinian groups. 
Note that they exist if and only if the boundary
surface has some discrete symmetry, and indeed we sit on a point in moduli
space with lots of symmetry. We have access to their actions since
they are simply related to  Fuchsian uniformization and the Liouville
action $S_L$  \cite{Yin:2007at}. It is believed that these solutions cannot be dominant
because they would lead to certain pathologies from the dual CFT
perspective - relating to studying the CFT on more than one disconnected
surface. But it remains an open question to show this.
\footnote{We thank Alex Maloney for discussion on this.} 
We are actually in a position to study this question however for now we leave
this to future work.

For the case $n=2$ one can  construct all the saddles. 
This is the case where $\mathcal{M}$ is a torus which is then
the same surface which is used to compute the thermal partition function of the CFT on a circle.
As discussed in \cite{Headrick:2010zt} the transition between the $\Gamma_\alpha$ and $\Gamma_\beta$ saddles  at $x=1/2$ is related to the black hole Hawking-Page transition.
There are also an infinite set of saddles  \cite{Maldacena:1998bw} which turn out  not to
contribute at large central charge for a purely imaginary torus modulus $\tau = i \tau_2$. 
These come from $SL(2,\mathbb{Z})$ modular transformations of the torus and
lead to the following saddle contributions to the Mutual Informatiom:
\be
\label{torussum}
I^{A,B}_2 (x) =  \frac{c \pi}{6} \frac{\tau_2}{ A^2 + B^2 \tau_2^2} - \frac{c}{12} \ln\left(2^8 (1-x)/x^2\right)
\ee
where $A,B \in \mathbb{Z}$ and are co-prime. The cycle
that is contractable in the bulk is $A C_{\alpha} + B C_{\beta}$
where $C_{\alpha, \beta}$ were defined in Figure~\ref{fig2cyc}.
Note that under the action of complex conjugation $(C_\alpha, C_\beta) \rightarrow (-C_\alpha,  C_\beta)$
so these bulk solutions break this symmetry when both $A$ and $B$ are not zero (
in which case there are multiple solutions with the same action.) 
However as is clear from \eqref{torussum} the dominant solutions
are either $(A=1,B=0)$ or $(B=1,A=0)$. We take this as evidence in
favor of the absence of replica symmetry breaking.

Finally note that understanding how the actions of these missing saddles 
can be continued to $n$ non-integer so we can take the limit $n \rightarrow 1$
is also important.  This seems rather tricky when the replica symmetry
is broken since we have to continue the symmetry braking pattern
to non-integer $n$ whatever that means. 

\subsection{Further work}

Aside from studying the possibility of replica symmetry breaking we see
several avenues for how to extend this work. Firstly it is important
to understand quantum corrections to the RT formula. This involves calculating
one loop determinants for fluctuations about the saddles that we constructed. 

Secondly one could try to generalize our setup in several ways. For example one could
try to work with different states in the CFT like finite temperature on a circle.
Moving to higher dimensions would be difficult because we no longer have
the power of $AdS_3$ holography. The power stems from the fact that
gravity has no propagating modes in three dimensions. Of course one could
still try to work numerically in higher dimensions, and as we have learned it
may not be necessary to solve the full numerical problem at integer $n$ in
order to reproduce RT.
It might be possible to proceed by simply setting up the problem well enough
so that in principle the ERE could be computed. Then if the bulk action
can be read off without reference to the specifics of the bulk solution it may
be possible to continue the answer to $n \rightarrow 1$ without doing
the hard numerical work.


\acknowledgements{We would like to thank Xi Dong, Tom Hartman,  Sean Hartnoll, Nabil Iqbal, 
Josh Lapan, Juan Maldacena, Alex Maloney and  Edward Witten for useful conversations. 
This work was supported by NSF grant PHY 0969448}

\appendix

\section{The Riemann surface}
\label{app:rsurf}

Specializing to two intervals $N=2$ the Riemann surface is defined by:
\be
\label{branch2}
y^n = \frac{ (z-z_1)(z-z_3)}{(z-z_2)(z-z_4)}
\ee

\begin{itemize}
\item $z$ and $y$ can be thought of meromorphic functions on the surface $\mathcal{M}$
\item $z$ defines a holomorphic map from $\mathcal{M}$ to $ \mathbb{C}{\rm P}^1$ branched at the points $z_i$.
The degree of the covering map is $n$. And the branching order is:
\begin{equation}
B_{z}(z_i) = n-1
\end{equation}
so that the Riemann-Hurwitz relation tells us that the genus of $M$ is $g = n (-1) + 1 + 2(n-1) = n-1$.
\item The surface has a cyclic $\mathbb{Z}_n$ symmetry (automorphisms of $M$) defined by cyclic permutations of the $n$ different sheets. It is generated by $(y \rightarrow y e^{i \pi/n}, \, z \rightarrow z)$. 
\item The space of complex structure deformations
of such a surface is $3(n-2)$ complex dimensional (for $n>2$), however 
 we are only interested in a one real dimension slice of this space. These
will be the deformations which preserve the $\mathbb{Z}_n$ rotations discussed above and
leave the surface in the form described by \eqref{branch2}.
\item A basis for the holomorphic differential forms is:
\be
\nu^t = \frac{d z}{y^t(z-z_2) (z - z_4)} \qquad t = 1, \ldots, n-1 
\ee
\item A linearly independent basis for the quadratic holomorphic differential forms is:\footnote{ In order to find these one notes that the point
defined by $z \rightarrow \infty$ is smooth 
and thus $\omega$  takes the form $P(z)dz^2 y^{-s}$ where $P(z)$ is meromorphic function with poles at the branch point and net pole order $-4$. There are linear relations that
need to be taken into account between these meromorphic functions. Then the bounds on $s$ can be found by examining the behavior close to each branch point. }
\bea
\omega^{s+1} & =&  \frac{d z^2}{y^s(z-z_1)(z-z_3)(z-z_2) (z - z_4)} \qquad s= 0, \ldots, n-2 \\
\omega^{s+n-2}  & =&  \frac{d z^2}{y^s(z-z_1)(z-z_3) (z-z_2)^2} \qquad s= 2, \ldots, n-2 \\
\omega^{s+2n-5}  & =&  \frac{d z^2}{y^s(z-z_1)(z-z_3) (z-z_4)^2} \qquad s= 2, \ldots, n-2 \\
\omega^{3(n-2)}  & =&  \frac{d z^2}{y^{n-1}(z-z_2)^2(z-z_4)^2} 
\eea
There are $3(n-2)$ of these, and they label the space of complex structure deformations.

\item Only one of the quadratic differentials is invariant under the $Z_n$ symmetry:
\be
\omega^{1}   \equiv  \frac{d z^2}{(z-z_1)(z-z_3)(z-z_2) (z - z_4)}
\ee

\end{itemize}

As usual we can use an $SL(2,\mathbb{R})$ transformation on $z$ to move the points $z_1,z_2,z_3,z_4$ to $0,x,1,\infty$ respectively.  After scaling $y$ by a constant the surface can be described by:
\begin{equation}
y^n = \frac{z (z-1)}{(z-x)} 
\end{equation}

\section{The connection matrices}
\label{app:conn}

In this section we setup the connection matrices which allow
us to study the monodromy problems numerically.
For each point $z_i$ consider the two linearly independent solutions:
\be
\label{ua1}
u^{(z_i)}(z) = \begin{pmatrix} \psi_+^{(z_i)}(z) \\ \psi_-^{(z_i)}(z) \end{pmatrix}
 \approx  \begin{pmatrix} (z-z_i)^{\frac{1}{2} + \frac{1}{2n}} \\  (z-z_i)^{\frac{1}{2}-\frac{1}{2n}}\end{pmatrix}
\ee
where $z$ is taken slightly to the right of $z_i$ on the real axis. 
Since we know the ode is real along the $z$ axis 
it is useful to construct the following real connection matrices:
\be
u^{(z_{i})}(z) =  R_{i,i+1} T u^{(z_{i+1})}(z)
\ee
where $T$ enacts the monodromy of moving $z$ from slightly to the right of $z_{i+1}$
to slightly to the left of $z_{i+1}$
\be
T = \begin{pmatrix}-  i e^{-i \pi/(2n)} & 0 \\ 0 & - i e^{i \pi/(2n)} \end{pmatrix}
\ee
From this it is clear that the remaining matrix $R_{i,i+1}$ is real.
Further the Wronskian condition on two
linearly independent solutions tells us that $\det R_{i,i+1} = -1$. 

To summarize we have defined the following set of $2N$ real connection problems:
\be \left.
\begin{aligned}
\psi &\approx \alpha_i (z-z_{i})^{\frac{1}{2} + \frac{1}{2n}}  + \beta_i (z-z_i)^{\frac{1}{2} - \frac{1}{2n}}\,
,\quad z \rightarrow z_i^+ \\
\psi & \approx    \alpha_{i+1} (z_{i+1}-z)^{\frac{1}{2} + \frac{1}{2n}}  + \beta_{i+1} (z_{i+1}-z)^{\frac{1}{2} - \frac{1}{2n}}\, ,\quad z \rightarrow z_{i+1}^- 
\end{aligned}  \quad
\right\}
\quad  \begin{pmatrix} \alpha_{i+1} & \beta_{i+1} \end{pmatrix}
= \begin{pmatrix} \alpha_i & \beta_i \end{pmatrix} R_{i,i+1}
\ee
which are amenable to numerical work. From these $R$'s and the matrix $T$ we can construct the full set of monodromies in a straightforward manner. 
For example the trivial monodromy condition on a path encircling two adjacent  points $(z_k, z_{k+1})$ is:
\be
{\mathbf 1} = T^2 R_{k,k+1} T^2 R_{k,k+1}^{-1}
\ee

\section{Regulating the twist operators}

\label{sec:cut}

The Liouville action \eqref{li} needs to be regulated by smoothing
out the singularities of $\phi_F$ close to the singular points $z_i$. 
Following \cite{Lunin:2000yv} we  cut out infinitesimal holes of $\mathcal M$ around these points $|z-z_i| < \epsilon$ and ``insert the unit operator'' by picking a different metric within these holes:
\be
ds^2 = d z d\bar{z} \left( |z-z_i|^2/\epsilon^2 \right)^{1/n-1}  \propto d t d \bar{t} \epsilon^{2-2/n}
\qquad |z-z_i| < \epsilon
\ee
and we have matched onto the metric $ds^2 = dz d\bar{z}$ 
at the boundary of the hole. Doing this allows
is to extend the definition of $\phi_F$ to the closed surface $\mathcal{M}$.
Note that for this choice of metric
 $\phi_F$ is essentially constant in this hole (up to corrections due to
the curvature of the hyperbolic space.) There are $4$ of these holes.

Additionally we also have to chop out $z > R_z$ which contributes
an IR divergence to the path integral.  
There are $n$ images of these $z \rightarrow \infty$ holes
in the $t$ plane at $t =  t_m^\infty$.
To do this we work with the following metric for large $z$:
\be
ds^2 =  R_z^4 \frac{ d z d \bar{z}}{|z|^4} \propto  d t d\bar{t} \,, \quad z > R_z \quad
(|t - t_m^\infty| < \# /R_z )
\ee
which again matches to $ds^2 = dz d\bar{z}$ at the boundary 
and sets $\phi_F$ to approximately a constant
in this region.  In this way the action $S_L$ in \eqref{li} is  well defined. It is easy to show that
the contributions to $S_F$ from the $n+4$ holes discussed above is zero essentially
because $\phi_F$ is a constant and the volume of the holes is going to zero in the limit
$\epsilon \rightarrow 0, R_z \rightarrow \infty$.
We can then restrict ourselves to performing the integral in
the Liouville action over the region
outside the holes: $\widehat{\mathcal{M}} = \mathcal{M}\backslash \{(|z-z_i|<\epsilon) \cup(|z|>R_z)\}$. 

We now write this integral in terms of the original $z$ coorindates:
\bea
\nonumber
S_L &=& \frac{c}{96 \pi} \int_{\widehat{\mathcal{M}}} \hspace{-.1cm} d^2 z 
\hspace{.2cm} \left( (\partial \phi_F)^2
+ 2 \phi_F ( \partial^2 \phi_F)\right) \\
&=& - \frac{c}{96 \pi}  \int_{\widehat{\mathcal{M}}} \hspace{-.1cm} d^2 z 
\hspace{.2cm} \left( (\partial \phi_F)^2 + 16 e^{-\phi_F}  \right)
 + S_{\mathrm bdry} - \frac{c}{3} (n-2)
 \label{liopol}
\eea
where we have integrated by parts and used the fact that the Euler character
of the surface $\mathcal{M}$ is $2(2-n)$ to introduce the Liouville term $\propto e^{-\phi_F}$
into the action. Up to this constant the final result is that of the standard Liouville action
for the field $\phi_F$ living on the $z$-plane. The boundary term also taking the
usual form:
\be
\left(\frac{96 \pi}{c} \right) S_{\mathrm bdry} = -  4\left(1-\frac{1}{n} \right) \int_{|z-z_i| =\epsilon} d \theta \phi_F
+ 8 \sum_{m=1}^{n} \int_{|z| = R_z} d\theta \phi_F
\ee
where the last sum is over the different sheets of the branched covering.
Also the first integral above is from $\theta =0, 2\pi n$.
We have used the behavior of $\phi_F$ close to $z_i$ and $z=\infty$:
\be
\label{bcl}
\phi_F \approx \left(1- \frac{1}{n}\right) \ln|z-z_i|^2 + \ldots \,, \quad z \rightarrow z_i\,;
\qquad \phi_F \approx 2 \ln |z|^2+ \ldots \,,\quad z \rightarrow \infty
\ee
This action actually defines the problem that we have already solved. That is
if we vary the action we find the Liouville equations of motion:
\be
\partial^2 \phi_F = - 8  e^{- \phi_F}
\ee
now subject to the boundary conditions \eqref{bcl}.
There is a unique solution to this problem. 
That is the solution which defines a constant
negative curvature on the space $\mathcal{M}$ and
which is equivalent to the solution found using the ode and Fuchsian monodromy condition.

\section{More on the symmetric domain action}
\label{sec:cut2}

We continue the calculation of the bulk gravity action starting from \eqref{inter}.
This action is integrated over the regulated domain $\widehat{D}_s$ defined
analogously to $\widehat{\mathcal{M}}$ in \eqref{liopol}.
That is $\widehat{D}_s = \widehat{D}_s \backslash \{ ( |z-z_i| < \epsilon) \cup
( |z| > R_z ) \cup (|w| > R_w ) \}.$

The field appearing in \eqref{inter} $\phi$ is the Weyl factor for
the the conformal transformation from the $w$ to $z$ planes:
\be
ds^2 = d z d\bar{z} = e^{-\phi} dw d\bar{w}
\ee
Thus  it will have singularities whenever either of these metrics is singular
(unless the two metrics have the same singularity and they cancels out.)
It is clear that $\phi$ is  harmonic so we can evaluate the bulk term of
\eqref{inter} by integrating by parts. This then reveals the singularities
located at $z_i$ and as well those at $z \rightarrow \infty$ which we remind
the reader are actually $n$ points one from each replica -  they are located on the $w$-plane at $w_\infty^m$. We also still have the $w$ plane singularity at $w  \rightarrow \infty$.
Altogether we have:
\bea
\nonumber
c^{-1} S_{gr} &=& \frac{1}{12} \phi( |w| = R_w ) -  \frac{1}{3} \ln R_w
+ \frac{1}{12} \sum_{m =1}^n \phi(|z|= R_z[w = w_\infty^m]) \\
&& - \frac{n-1}{24} \sum_{i=1}^4 \phi(|z - z_i| = \epsilon) 
- \frac{n}{24 \pi} (I_1 -8 N_1) + \frac{(n-2)}{3} \ln \Lambda  + \frac{(n-2)}{3}
\label{sings}
\eea
where we have defined the integral:
\be
\label{intw}
I_1 = \int_{U_1} d \theta \left( \phi - 4 \ln \rho\right)
\ee
and the notation is for this integral the same as in \eqref{volume1}.

In order to make sense of \eqref{sings} we have
temporarily deformed the space $w$ such
that $z_4$ no longer maps to $ w = \infty$ but instead maps to $w_4 \gg 1$ and also
such that there is a point $z_\infty \approx z_4$ (on one of the replicas) which does map to 
to $w =\infty$. This will disentangle the $w$-plane curvature which
is hidden at $w \rightarrow \infty$ from the twist operator at $z=z_4$ which
previously mapped to $w = \infty$ (see Figure~\ref{fig:zw}). The behavior of $\phi$ can then
be read of from the behavior of the Schottky uniformization coordinate $w(z)$:
\begin{itemize}
\item Around the twist operators $z \rightarrow z_i$:
\be
w(z) \approx w_i  + \mu_i (z-z_i)^{1/n}\,\, \rightarrow \,\, \phi \approx 2 \left( \frac{1}{n} -1\right)
\ln|z-z_i| + 2 \ln |\mu_i/n| 
\ee
\item Around the $w$-plane curvature singularity at $w\rightarrow \infty$ (recall
that $z_\infty$ is close to $z_4$ and only on one of the replicas):
\be
w(z) \approx \frac{ \nu_\infty}{ z-z_\infty}  \,\, \rightarrow \,\, \phi \approx 4 \ln |w| - 2 \ln  |\nu_\infty|
\ee
\item Around the  $z$-plane curvature singularities at $z \rightarrow \infty$
\be
w(z) \approx w_{\infty}^m + \frac{ \mu_{\infty}^m}{z} \,\,\rightarrow
\,\, \phi \approx - 4 \ln |z| + 2 \ln | \mu_\infty | \qquad ( m = 1,\ldots n)
\ee
\end{itemize}
where we have defined many new constants $\mu_i , \mu_\infty^m, \nu_\infty$ 
which can be extracted from $w(z)$ numerical.
Plugging this into \eqref{sings} we find:
\bea
c^{-1} S_{gr}
 &=& - \frac{(n-1)}{12} \sum_{i=1}^4 \ln| \mu_i/n|  - \frac{1}{6}  \ln |\nu_\infty| 
+ \frac{1}{6} \sum_{m=0}^{n-1} \ln | \mu^m_\infty| 
\\
\nonumber
&&  \hspace{.2cm} + \frac{(n-2)}{3} - \frac{n}{24 \pi} ( I_1 - 8 N_1)  - \frac{ n}{3} \ln R_z - \frac{(n-2)}{3}\ln \Lambda
+\frac{ (n-1)^2}{3 n} \ln \epsilon
\label{nus}   
\eea
Note that $|\mu_\infty^m|$ will actually all be the same by the replica symmetry.
We now want to return to  the setup of interest  (after the aforementioned
deformation of the $w$ plane) 
by taking the limit $w(z_4) \rightarrow \infty$ and thus $z_\infty \rightarrow z_4$.
We also want to eventually take the limit $z_4 \rightarrow \infty$
(and set $z_1 = 0, z_2 = x, z_3 = 1$) since this is most convenient for
numerical work. We take these limits sequentially:
\begin{itemize}
\item In the coincidence limit close to $w \approx w_4 \approx \infty$ 
one can argue that the analytic map has the form:
\be
\label{largew4}
z \approx z_4 + w_4^{2n} \mu_4^{-n} \left( w_4^{-1} - w^{-1} \right)^n
\ee
which reproduces the behavior about $w =w_4$ and $w = \infty$ if
we additionally have $\nu_\infty = - n w_4^{n+1}/\mu_4^{n} $.  Expanding
\eqref{largew4} for large $w_4$ we have:
\be
w \approx \widetilde{\mu}_4 (z-z_4)^{-1/n}\,, \qquad \widetilde{\mu}_4
 = - w_4^2 / \mu_4
\ee
where we should fix $\widetilde{\mu_4}$ in this limit. The we find:
\be
\lim_{w_4 \rightarrow \infty } | \nu_{\infty}^2 \mu_4^{n-1} |
= | \widetilde{\mu}_4|^{(n+1)} n^{2(n-2)}
\ee

\item Similarly in the limit $z_4 \rightarrow \infty$ we have:
\be
\lim_{z_4 \rightarrow \infty } |(\mu_{\infty}^1)^{2n} \mu_4^{-n-1}|
= |\widehat{\mu}_4^{n-1} z_4^{2(n-1/n)}| n^{-2/n}
\ee
where we have defined as $z \rightarrow \infty$:
\be
w \approx \widehat{\mu}_4 z^{1/n}
\ee
\end{itemize}
Taking these limit in \eqref{nus} gives:
\bea
c^{-1} S_{gr}
 &=& 
 -  \frac{(n-1)}{12} \ln \left| \frac{\mu_1 \mu_2 \mu_3}{\widehat{\mu}_4} \right| 
- \frac{n}{24 \pi} (I_1 - 8 N_1) + \frac{(n-3)}{6} \ln(n) + \frac{(n-2)}{3} \\
& & + \frac{ (n^2-1)}{6 n} \ln z_4  + \frac{ (n-1)^2}{3 n} \ln \epsilon -  \frac{n}{3} \ln R_z
- \frac{(n-2)}{3} \ln \Lambda 
\nonumber
\eea
To deal with all the regulator factors
we simply compute the mutual information. We need
the ERE for a single interval using the same technique as above, so that
the non-universal pieces cancel. One finds (the uniformization
map can be found in this case analytically, we do not
go through the details which can be for example found in
\cite{Lunin:2000yv}):
\be
c^{-1} S_{gr}^{N=1} = \frac{1}{6} \left( n - \frac{1}{n} \right)
\ln | z_1 - z_2 | - \frac{1}{3} \ln n - \frac{1}{3} - \frac{n}{3} \ln R_z
+ \frac{1}{3} \ln \Lambda
\ee
where $z_1, z_2$ are the end points of the single interval.
We also need the partition function $Z_1$
for the theory on a single replica 
in order to relate $Z_\mathcal{M}$ to the EE. See \eqref{renyicomp}.  This can be computed as above to find $\ln Z_1 = c/3( \ln(R_z/\Lambda) + 1)$. 
Putting everything together we find:
\be
 c^{-1} I_n^\beta = 
\frac{1}{12} \ln \left| \frac{\mu_1 \mu_2 \mu_3}{\widehat{\mu}_4} \right|
 + \frac{n}{24 \pi(n-1)} (I_1 - 8N_1)  + \frac{1}{6} \left( \frac{1}{n} + 1 \right)
\ln(x)    + \frac{ (n+1)\ln(n)}{6 (n-1)}  
\ee
where we remind the reader this computation was for the $\Gamma_\beta$
saddle. After mapping the integral $I_1$ in \eqref{intw} to the $z$ plane
we get the expression quoted in \eqref{master}.

\section{Quasi-conformal transformations}

\label{sec:quasi}

A variation of the modular parameter of the surface
$\mathcal{M}$ can be thought of as a quasiconformal transformation. 
A transformation which deforms infinitesimally the complex structure of the manifold $\mathcal{M}$. It was 
shown in \cite{zta} that the ZT action behaves nicely under quasiconformal transformations.
The results make physical sense since the variation gives an integral
of the stress tensor associated with the Schottky uniformization:
\be
\label{ztvar}
\delta S_{ZT} = \frac{c }{2 4 \pi} \int_{\mathcal{M}} d^2 w \left( \delta N_{\,\,\bar w}^w \hat{T}_{ww}
+\delta \bar{N}_{\,\, w}^{\bar{w}} \bar{\hat{T}}_{\bar{w}\bar{w}} \right)
\qquad \hat{T}_{ww} = \frac{1}{2} (\partial_w \phi_S)^2
+ \partial_w \partial_w \phi_S   
\ee
The quasi-conformal variation is defined through a Beltrami differential
$\delta N_{\,\,\bar w}^w$.To describe this consider varying the complex structure 
of $\mathcal{M}$ by  defining  new holomorphic coordinates on the different
coordinate patches. For example working on the branched covering:
\be
z' = z +\delta z (z,\bar{z})  \qquad y'=  y+  \delta y (y,\bar{y}) 
\ee
and equivalently for the antiholomorphic coordinates. 
Demanding that the new transition functions (defined by the
complex curve \eqref{branch2}) are holomorphic with respect
to the coordinates $(z',y', \ldots)$ defines a Beltrami differential:
\be
\delta N_{\,\,\bar z}^z = \partial_{\bar z} \delta z(z,\bar{z})
\ee
which transforms as the placement of the indices suggests. It is this type of
differential that appears in \eqref{ztvar}.  The integral in \eqref{ztvar} is simply over the entire surface $\mathcal{M}$ without boundary (compared to the integrals
over the fundamental $\mathcal{D}_d$) since the action of $\Sigma$ does not change the stress tensor $\hat{T}$ and $\delta N$ is well defined on the whole surface.

Let us transform this
result to the $z$ plane:
\begin{align}
\delta S_{ZT} &= \frac{c }{2 4 \pi} \int_{\mathcal{M}} d^2 z\left(   \hat{T} _{zz}  \partial_{\bar z} \delta z 
 \vphantom{\lim_{a}} +   \bar{\hat{T}}_{\bar z \bar z}  \partial_{z} \delta \bar{z} 
  \right) \qquad
\hat{T}_{zz} =   \{ w,z\} - \{ t,z \} \\
&=   \frac{c }{2 4 \pi} \sum_i  \int_{\mathcal{M}} d^2 z \left( \partial_{\bar z} \delta z(z,\bar{z})
\frac{ ( p_i - p_i^F)}{(z-z_i)} + \partial_{ z} \delta \bar{z}(\bar{z},z)
\frac{ ( \bar{p}_i -\bar{p}_i^F)}{(\bar{z}-\bar{z}_i)} \right)
\label{invert}
\end{align}
where we have replaced the the Schwarzian derivatives
with the their expansion in terms of accessory parameters \eqref{stress}.  
In particular we have set all replica symmetry breaking
accessory parameters to zero. We have picked $\delta z$ to be
the same on each replica and in this way the quasiconformal
transformations leaves us on the moduli space of Riemann surfaces
defined by \eqref{branch2}. 

Note that what we have arrived at in \eqref{invert}  looks like a Ward identity for conformal invariance. Integrating we find,
\be
\delta S_{ZT} =\frac{ cn}{12} \sum_i ( (p_i - p_i^F) \delta z_i + c.c. )
\ee
where $\delta z_i \equiv \delta z(z_i,\bar{z_i})$.
We have integrated over all the replicas explaining the factor of $n$.
A certain amount of regularity in $\delta z(z,\bar{z})$ was assumed
in order to be able to invert the operator $\partial_{\bar z}$ in \eqref{invert} and this
can be justified by giving an explicit expression for $\delta z$
in terms of $\delta z_i$
\be
\delta z(z,\bar{z}) =  - \frac{1}{2} e^{\phi_F} \sum_{i} \delta z_i \partial_{z_i} \partial_{\bar{z}} \phi_F
\ee
where $\phi_F$ is the Fuchsian Liouville field and this
equation is the same on each replica. See \cite{ztc} for the complete discussion.

\end{document}